%% file: main.tex
\documentclass[preprint,3p, 12pt]{elsarticle}

\usepackage{amssymb}
\usepackage{amsthm}

\theoremstyle{definition}
\newtheorem{theorem}{Theorem}[section]
\newtheorem{example}[theorem]{Example}
\newtheorem{proposition}[theorem]{Proposition}
\newtheorem{lemma}[theorem]{Lemma}
\newtheorem{definition}[theorem]{Definition}
\newtheorem{remark}[theorem]{Remark}

\newtheorem{corollary}[theorem]{Corollary}

\usepackage{algorithm}
\usepackage{algpseudocode}

\usepackage{enumitem}

\usepackage{mathrsfs}

\usepackage{amsmath}
\usepackage{xcolor}

\usepackage{preamble}

\journal{Journal of Symbolic Computation}

\begin{document}

\begin{frontmatter}



\title{Further results on Minimal and Minimum Cylindrical Algebraic Decompositions}


\author[inst1]{Lucas Michel}
\ead{lucas.michel@uliege.be}
\affiliation[inst1]{organization={University of Liege, Mathematics research unit},
            addressline={Allée de la découverte, 12}, 
            city={Liège},
            postcode={B-4000},
            country={Belgium}}

\author[inst1]{Pierre Mathonet}
\ead{p.mathonet@uliege.be}
\author[inst1]{Naïm Zénaïdi}
\ead{nzenaidi@uliege.be}


\begin{abstract}
We consider cylindrical algebraic decompositions (CADs) as a tool for representing semi-algebraic subsets of $\mathbb{R}^n$. In this framework, a CAD $\mathscr{C}$ is adapted to a given set $S$ if $S$ is a union of cells of $\mathscr{C}$. 
Different algorithms computing an adapted CAD may produce different outputs, usually with redundant cell divisions. In this paper we analyse the possibility to remove the superfluous data.  We thus consider the set $\CAD^r(\Fr)$ of CADs of class $C^r$ ($r \in \N \cup \{\infty, \omega\}$) that are adapted to a finite family  $\mathcal{F}$ of semi-algebraic sets of $\R^n$, endowed with the refinement partial order and we study the existence of minimal and minimum element in $\CAD^r(\Fr)$. We show that for every such $\Fr$ and every $\mathscr{C} \in \CAD^r(\Fr)$, there is a minimal $\CAD$ of class $C^r$ adapted to $\Fr$ and smaller (i.e. coarser) than or equal to $\mathscr{C}$. In dimension $n=1$ or $n=2$, this result is strengthened by proving  the existence of a minimum element in $\CAD^r(\Fr)$. In contrast, for any $n \geq 3$, we provide explicit examples of semi-algebraic sets whose associated poset of adapted CADs does not admit a minimum. We then introduce a reduction relation on $\CAD^r(\Fr)$ in order to define an algorithm for the computation of minimal CADs and we characterise those semi-algebraic sets $\Fr$ for which $\CAD^r(\Fr)$ has a minimum by means of confluence of the associated reduction system.
We finally provide practical criteria for deciding if a semi-algebraic set does admit a minimum CAD and apply them to describe various concrete examples of semi-algebraic sets, along with their minimum $\CAD$ of class $C^r$. 
\end{abstract}



\begin{keyword}
Semi-algebraic set \sep cylindrical algebraic decomposition \sep partially ordered set \sep minimal and minimum element \sep abstract reduction system
\MSC 14P10
\end{keyword}

\end{frontmatter}



\input{introduction}
\input{preliminaries}
\input{minimal}
\input{minimum}

\input{reductions}
\input{positive}
\input{behaviour}

\section*{Acknowledgments}
This article is a revised and expanded version of \cite{miniCAD}. The authors would like to express their gratitude to J. H. Davenport for his advice during the elaboration of this work. It is a pleasure to thank B. Boigelot, M. England and P. Fontaine for fruitful discussions. 
  P.~Mathonet, L.~Michel and N.~Zenaïdi are supported by the FNRS-DFG PDR Weaves (SMT-ART) grant 40019202.






 \nocite{*} 
 \noindent
 \bibliographystyle{elsarticle-num} 
 \noindent
 \bibliography{references}






\end{document}

%% file: introduction.tex
\section{Introduction}

Cylindrical algebraic decomposition (CAD) is a fundamental tool in computer algebra and in real algebraic geometry. Since its introduction by Collins \cite{collins1975} in the framework of real quantifier elimination, the scope of applications of CAD and the concept itself have evolved in various directions. Algorithms computing CADs are constantly refined \cite{LazardStyle23, McCallum98, Brown01, RegularChains} and are nowadays used as building blocks in various fields where symbolic manipulations of polynomial equalities and inequalities are relevant.

From a theoretical standpoint, the existence of a CAD algorithm may also be seen as a structure theorem for semi-algebraic sets not only giving topological information about them (see for example \cite{bochnaketal1998, schwartzsharir83,basu2007}) but also providing a constructive and explicit representation of these subsets which enables symbolic and effective reasoning. In this framework, we say that a CAD is adapted (or represents) a semi-algebraic set $S \subseteq \mathbb{R}^n$ if $S$ is a union of cells from this CAD.

In practice, for computing a CAD adapted to a semi-algebraic set $S$, one describes $S$ by polynomial inequalities (conjunction, disjunction, and negation) associated with a finite set of $n$-variate polynomials $\mathcal{P}$. 
The CAD algorithms usually take $\mathcal{P}$ as input and the output CAD is therefore adapted to any semi-algebraic set defined by means of $\mathcal{P}$. For this reason they usually contain unnecessary cells for the description of $S$. Moreover, these outputs also depend on the choice of $\mathcal{P}$ and on the considered algorithm. It is then natural to seek for a method for simplifying a given CAD, while retaining only the relevant information needed to describe $S$ (see for instance \cite{BrownSimple}). 

The analysis of CADs adapted to a single semi-algebraic set naturally leads us to CADs that are simultaneously adapted to a finite family of sets. This situation is reminiscent of what happens with CAD algorithms when $\mathcal{P}$ contains one polynomial. We thus consider a finite family $\F$ of semi-algebraic sets of $\R^n$ and study in detail the set $\CAD^r(\F)$ of CADs (with respect to a fixed variable ordering) of class $C^r$ ($r \in \N \cup \{\infty, \omega\}$) that are adapted to every element of $\F$. 
This set is naturally endowed with the partial order $\preceq$ defined by refinement: for $\mathscr{C}, \mathscr{C}' \in \text{CAD}^r(\F)$, we write $\mathscr{C} \preceq \mathscr{C}'$ if every cell of $\mathscr{C}$ is a union of cells of $\mathscr{C}'$. Here, the main problems\footnote{Particular instances of these problems were already mentioned in \cite{wilson,locatelli}.} under consideration are the existence and the effective construction of minimal and minimum elements of the poset $(\text{CAD}^r(\F), \preceq)$.
These problems, theoretical in nature, lead us to analyse in detail the question of simplifying CADs by merging cells (see Section \ref{sec:red}).
From a more practical perspective, the minimum number of cells of these minimal CADs provides a tight lower bound on the number of cells that a CAD adapted to a semi-algebraic set must contain and could therefore be used to benchmark CAD algorithms (see for example \cite{NuCAD, MLCAD}).  


In Section \ref{sec:minimal}, we develop the first results on minimal CADs. We show in particular that for every finite family $\Fr$ of semi-algebraic sets of $\mathbb{R}^n$ and every CAD $\mathscr{C}\in\CAD^r(\F)$, there is a minimal CAD adapted to $\F$ and smaller (i.e. coarser) than or equal to $\mathscr{C}$. This elementary result settles the question of the existence of minimal CADs adapted to $\F$.  
Section \ref{sec:minimum} initiates the study of minimum CADs. More specifically, if $n=1$ or $n=2$, we prove the existence of a minimum element in $\CAD^r(\F)$ for every finite family $\Fr$ of semi-algebraic sets of $\mathbb{R}^n$. We actually use the results concerning minimal CADs obtained in Section \ref{sec:minimal} to show that such CADs must coincide. On the contrary, for $n \geq 3$, we show that there exist semi-algebraic sets whose associated poset of adapted CADs of class $C^r$ does not admit a minimum by providing explicit examples. 
The previous positive results in dimension $n=1$ and $n=2$, together with the counterexamples in any higher dimension, lead naturally to the investigation of those semi-algebraic sets admitting a minimum CAD, together with the development of effective tools to determine such CADs.
For this purpose, in Section \ref{sec:red}, we study a reduction relation on the set $\text{CAD}^r(\F)$. We first use it to define an algorithm that constructs a minimal CAD adapted to a given family $\F$ and smaller than or equal to a given CAD $\mathscr{C}$. This algorithm may be used as a post-processing operation for any CAD algorithm. 
We then employ this relation to characterise, in terms of the confluence of the associated reduction system, those finite families $\Fr$ of semi-algebraic sets for which there exists a minimum adapted $\CAD$ of class $C^r$.
In Section \ref{sec:positive}, we prove an inductive criterion stating that in general $\CAD^r(\Fr)$ admits a minimum if and only if a certain collection $\pi_{n-1}(\CAD^r(\Fr))$ of CADs of $\R^{n-1}$ does. We elaborate on this criterion by considering overapproximations $\mathcal{O} \supseteq \pi_{n-1}(\CAD^r(\Fr))$ that enable us to obtain sufficient conditions for the existence of a minimum in $\CAD^r(\Fr)$. As an application, we apply this machinery to describe explicitly minimum CADs adapted to various semi-algebraic sets. 


%% file: preliminaries.tex
\section{Background and notation}
We fix a positive integer $n \in \mathbb{N}^*$ and an element $r \in \N \cup \{\infty, \omega\}$. Recall that the order on $\N$ is usually extended by defining $s < \infty < \omega$ for all $s \in \N.$ 

We consider CADs of $\mathbb{R}^n$ of class $C^r$ with respect to a fixed variable ordering. Since the cells of such CADs are defined inductively from cells of CADs of $\mathbb{R}^k$ ($k \leq n$) which are indexed by $k$-tuples, it is convenient to make tuple notation more concise by identifying the $k$-tuple $I=(i_1,\ldots,i_k)\in\mathbb{N}^k$ with the corresponding word $i_1\ldots i_k$. We say that $I$ is odd (resp. even) if $i_k$ is odd (resp. even). We denote by $\varepsilon$ the empty tuple, which corresponds to the empty word. For $j\in \mathbb{N}$, we denote by $I:j$ the $k+1$ tuple $(i_1,\ldots,i_k,j)$.

\begin{definition}[see \cite{Arnon,basu2007}]\label{def:cad}
    A cylindrical algebraic decomposition ($\CAD^r$) of $\mathbb{R}^n$ of class $C^r$ is a sequence $\mathscr{C} = (\mathscr{C}_1,\ldots,\mathscr{C}_n)$ such that  
    for all $k \in \{1,\ldots,n\}$, the set  $$\mathscr{C}_k= \left\{C_{i_1 \cdots i_k} |  \forall j \in \{1,\ldots,k\}, i_j \in \{1,\ldots,2u_{i_1\cdots i_{j-1}} +1\}\right\}$$
     is a finite semi-algebraic partition of $\mathbb{R}^k$  defined inductively by the following data:
    \begin{enumerate}
        \item there exists a natural number $u_\varepsilon \in \mathbb{N}$ and real algebraic numbers\footnote{\label{note:none}Possibly none if $u_\varepsilon = 0$.
        }
        $\xi_{2} < \xi_{4} < \ldots <\xi_{2u_\varepsilon}$
        that define exactly all cells of $\mathscr{C}_{1}$ by
        \[C_{2j} = \{\xi_{2j}\},\, (1 \leq j \leq u_\varepsilon),\quad
             C_{2j+1}= (\xi_{2j}, \xi_{2(j+1)}),\,(0 \leq j \leq u_\varepsilon)
         \]
         with the convention that $\xi_{0} = -\infty$ and $\xi_{2u_\varepsilon+ 2} = +\infty$;
        \item for each cell $C_I \in \mathscr{C}_k$ ($k < n$), there exists a natural number $u_I \in \mathbb{N}$ and semi-algebraic functions\footnote{Possibly none if $u_I = 0$. Recall that $\xi_{I:2j} < \xi_{I:2(j+1)}$ means that $\xi_{I:2j}(x) < \xi_{I:2(j+1)}(x)$ for all $x \in C_I$.} 
         $\xi_{I: 2 } < \xi_{I:4} < \ldots <\xi_{I:2 u_I} : C_I \to \mathbb{R}$ of class $C^r$, 
        that define exactly all cells of $\mathscr{C}_{k+1}$ by
        {\small\begin{align*}
             C_{I:2j} &= \{(\textbf{a}, b) \in C_I \times \mathbb{R} \; | \; b = \xi_{I:2j}(\textbf{a})\} , \quad (1 \leq j \leq u_I),\\
             C_{I:2j+1} &= \{(\textbf{a}, b) \in C_I \times \mathbb{R} \; |
\;\xi_{I:2j}(\textbf{a}) < b < \xi_{I:2(j+1)}(\textbf{a})\},\quad (0 \leq j \leq u_I)
         \end{align*}}
         with the convention that $\xi_{I:0} = -\infty$ and $\xi_{I:2u_I 
+ 2} = +\infty$.
    \end{enumerate}
 We say that the element $C_{I}$ of $\mathscr{C}_k$ is a cell of index $I$. Notice that it is either a singleton or an embedded submanifold of $\R^{k}$ of class $C^r$ diffeomorphic to an open ball of some dimension less than or equal to $k$. If $I$ is odd (resp. even), we say that this cell is a sector (resp. a section). When the context is clear, we identify $\mathscr{C}$ with $\mathscr{C}_n$.

\end{definition}
We usually denote by $S$ a semi-algebraic set of $\mathbb{R}^n$ and we denote by $S^c$ its complement in $\mathbb{R}^n$.
We are mainly interested in CADs that are adapted to $S$.  However, it will also be useful to consider CADs that are simultaneously adapted to a finite family of semi-algebraic sets. As we continue, we usually denote by $\mathcal{F}$ a finite family $\left(S_1,\ldots,S_p\right)$ of semi-algebraic sets in $\R^n$, with $p \in \N$ and we set $\Fr^c = \left(S_1^c, \ldots, S_p^c\right)$.  

\begin{definition}\label{def:adaptedF}
    We say that a CAD $\mathscr{C}$ is adapted to $S$ if $S$ is a union of cells of $\mathscr{C}$. We denote by $\text{CAD}^r(S)$ the set of all CADs of class $C^r$ which are adapted to~$S$.
    More generally, we say that a CAD $\mathscr{C}$ is adapted to $\mathcal{F}$ if $\mathscr{C}$ is adapted to $S_i$ for every $i\in\{1,\ldots,p\}$. We denote by $\text{CAD}^r(\mathcal{F})$ the set of all CADs of class $C^r$ adapted to~$\mathcal{F}$. \red{}
\end{definition}

Observe that by definition, we have $\text{CAD}^r(\mathcal{F}) = \text{CAD}^r(S_1) \cap \ldots \cap \text{CAD}^r(S_p)$. Moreover,  for all $r, s \in \N \cup \{\infty, \omega\}$ such that $r<s$, we have $\CAD^r(\Fr) = \CAD^s(\Fr)$ if $n=1$  but $\CAD^s(\mathcal{F})\subset\CAD^r(\mathcal{F})$ if $n > 1$. 
\begin{remark}\label{rem:Collins} 
    The set $\text{CAD}^r(\mathcal{F})$ is never empty (see for instance Section 4 of \cite{vDDM}). Historically, Collins' seminal theorem \cite{collins1975} asserts that $\text{CAD}^0(\mathcal{F})$ is never empty. In particular, if we know the finite set $\mathcal{P}$ of polynomials involved in a description of a semi-algebraic set, then we can construct algorithmically a CAD of class $C^0$ adapted to $\Fr$. Under some suitable assumptions on $\mathcal{P}$, it is possible to build an element of $\CAD^\omega(\mathcal{F})$ (see~\cite{McCallum98}).
\end{remark}

The following lemma rephrases Definition \ref{def:adaptedF}.

\begin{lemma}\label{lemma:compl-proj}
        A CAD $\mathscr{C}$ is adapted to $\Fr$ if and only if for every $C\in \mathscr{C}$ and $i \in \{1, \ldots, p\}$ such that $C\cap S_i\neq\varnothing$, we have $C\subseteq S_i$. In particular, we have $\CAD^r(\Fr) = \CAD^r(\Fr^c).$ 
\end{lemma}
\begin{proof}

Consider a CAD $\mathscr{C}$ adapted to $\Fr$, $C \in \Cr$ and $x \in C \cap S_i$. Since $S_i$ is a union of cells of $\Cr$, there exists a cell $C'\in\Cr$ such that $x\in C'\subseteq S_i$. Then $C'\cap C\neq\varnothing$,  hence $C=C'\subseteq S_i$. Conversely, suppose that every cell that intersects $S_i$ lies entirely in $S_i$, and consider $x\in S_i$. Then, there is a cell $C \in \Cr$ such that $x\in C$. By hypothesis, we have $C\subseteq S_i$, so $S_i$ is a union of cells of $\Cr$.
If $\Cr \in \CAD^r(\Fr)$, then for $i\in\{1,\ldots,p\}$ and $x\in S_i^c$ the cell $C\in\Cr$ containing $x$ is a subset of $S_i^c$, since otherwise we would have $C\subseteq S_i$, which is impossible. Therefore $S_i^c$ is a union of cells of $\Cr$ and $\CAD^r(\Fr) \subseteq \CAD^r(\Fr^c)$. The other inclusion follows by symmetry.
\end{proof}

Since CADs are partitions, we naturally use the refinement order defined on the set of partitions to compare them.  

\begin{definition}\label{def:order}
    Let  $\mathscr{C}$ and $\mathscr{C}'$ be two CADs (or more generally partitions) of $\mathbb{R}^n$. We say that $\mathscr{C}'$ is a refinement of $\mathscr{C}$ if every cell of $\mathscr{C}$ is a union of cells of $\mathscr{C}'$. We also say that $\mathscr{C}$ is smaller than or equal to $\mathscr{C}'$ and write $\mathscr{C} \preceq \mathscr{C}'$.  We write $\mathscr{C} \prec \mathscr{C}'$ if $\mathscr{C} \preceq \mathscr{C}'$ and $\mathscr{C} \neq \mathscr{C}'$.
    \end{definition}
    This definition is already used in \cite{BrownSimple}, where $\mathscr{C}$ is said to be simpler than $\mathscr{C}'$ when $\mathscr{C} \prec \mathscr{C}'$ and where algorithms to compute simple  CADs with respect to this order are devised.
We also recall the definition of minimal and minimum elements. 
    \begin{definition}
    A minimal $\CAD^r$ adapted to $\mathcal{F}$ is a minimal element of ($\text{CAD}^r(\mathcal{F}),\preceq$). Namely, $\mathscr{C} \in \text{CAD}^r(\mathcal{F})$ is a minimal $\CAD^r$ adapted to $\mathcal{F}$ if 
        \[\forall \mathscr{C}' \in \text{CAD}^r(\mathcal{F}), (\mathscr{C}'\preceq \mathscr{C})  \implies (\mathscr{C}' =  \mathscr{C}).\] 
        A minimum $\CAD^r$ adapted to $\mathcal{F}$ is a minimum element of ($\text{CAD}^r(\mathcal{F}),\preceq$). Namely,  $\mathscr{C} \in \text{CAD}^r(\mathcal{F})$ is a minimum $\CAD^r$ adapted to $\mathcal{F}$ if 
    \[\forall \mathscr{C}' \in \text{CAD}^r(\mathcal{F}), \mathscr{C}\preceq\mathscr{C}'.\] 
        We say that $\mathcal{F}$ admits a minimum $\CAD^r$ if there exists a minimum $\CAD^r$ adapted to $\mathcal{F}$. 
    \end{definition}


%% file: minimal.tex
\section{Minimal CAD{\small s}}\label{sec:minimal}
In this section, we show that for every finite family $\Fr$ of semi-algebraic sets of $\R^n$, the poset $(\CAD^r(\Fr), \preceq)$ admits at least a minimal element. Then we show the equivalence between the uniqueness of a minimal $\CAD^r$ and the existence of a minimum $\CAD^r$. Finally, we give uniqueness results concerning the set of sections of the last level of minimal CADs. 

\begin{proposition}\label{prop:existMin}
If $\mathscr{C}$ is a $\CAD^r$ adapted to $\F$, then there exists a minimal $\CAD^r$ adapted to $\F$ that is smaller than or equal to $\mathscr{C}$. 
\end{proposition}

\begin{proof}
The set  $\text{SSP}(\mathscr{C})$ of partitions that are strictly smaller than $\mathscr{C}$ is a finite poset, by definition. We can find a minimal element by inspection: if $\mathscr{C}$ is not a minimal element in $\text{CAD}^r(\F)$, we can find $\mathscr{C}'$ in $\text{CAD}^r(\F)$ such that $\mathscr{C}'\prec \mathscr{C}$, and iterate the procedure with $\mathscr{C}'$.  This will end in a finite number of steps because $\text{SSP}(\mathscr{C})$ is finite and strictly contains $\text{SSP}(\mathscr{C}')$. Furthermore, this procedure will obviously result in a minimal CAD with the required properties. 
\end{proof}

\begin{remark}\label{rem:naive}
   The determination of a minimal element in $\text{CAD}^r(\F)$ presented in the proof of Proposition \ref{prop:existMin}  is in general not efficient. If $\mathscr{C}$ contains $K$ cells, then the set SSP$(\mathscr{C})$ of partitions that are strictly smaller than $\mathscr{C}$ contains $B_K - 1$ elements, where $B_K$ is the $K^\text{th}$ Bell number. This procedure will be improved by Algorithm \ref{algo:Min}.
\end{remark}

\begin{proposition}\label{prop:uniqueMin}
    There is a unique minimal $\CAD^r$ adapted to $\F$ if and only if there exists a minimum $\CAD^r$ adapted to $\F$.
\end{proposition}
\begin{proof}
    Let $\mathscr{M}$  be the unique minimal $\CAD^r$ adapted to $\Fr$. If $\mathscr{C} \in \text{CAD}^r(\Fr)$, Proposition \ref{prop:existMin} gives us a minimal $\CAD^r$ adapted to $\Fr$ and smaller than or equal to $\mathscr{C}$. By assumption, this minimal $\CAD^r$ is $\mathscr{M}$. This shows that the CAD $\mathscr{M}$ is smaller than or equal to any $\CAD^r$ adapted to $\F$, hence $\mathscr{M}$ is a minimum of $\text{CAD}^r(\F)$. The other implication is direct.
\end{proof}

In view of Proposition \ref{prop:uniqueMin} it is important to study further the properties of minimal CADs adapted to a family $\F$ in order to decide whether or not they must coincide. The following lemma provides a partial result in this direction concerning the last level of such minimal CADs.  In order to have the same treatment for the general case and for the special case of dimension one, it is useful to consider the real algebraic numbers $\xi_{2}, \ldots, \xi_{2u_\varepsilon}$ that define a CAD of $\mathbb{R}$ as images of semi-algebraic functions defined on the one point set $\mathbb{R}^0\cong\{0\}$, which we still denote $\xi_{2j}$, that is, we set $\xi_{2j}=\xi_{2j}(0)$. It will also be convenient to consider the cells of a CAD $\mathscr{C}_1$ of $\mathbb{R}$ to be built above the set $\mathbb{R}^0$ endowed with the CAD $\mathscr{C}_0=\{C_{\varepsilon}\}$ with $C_\varepsilon=\{0\}$.
For every $x\in\mathbb{R}^{n-1}$, and every subset $S$ of $\R^n$ we define $S_x= \{y \in \mathbb{R} \; |\; (x,y) \in S\}\subseteq \mathbb{R}$  and denote by $\partial S_x$ its topological boundary in $\mathbb{R}$. 

\begin{proposition}\label{rem:top-caract}
If $\mathscr{C} \in \text{CAD}^r(\F)$, $C_I \in \mathscr{C}_{n-1}$ and $x \in C_I$, then $$\cup_{i=1}^p\partial (S_{i})_x\subseteq \{\xi_{I:2j}(x)  \; | \; j \in \{1,\ldots,u_I\}\}.$$ If $\mathscr{C}$ is minimal, these two sets are equal. 
\end{proposition}
\begin{proof}
Consider $i\in\{1,\ldots,p\}$ and $y\in\partial (S_i)_x$ and suppose that $y$ is not in $\{\xi_{I:2j}(x) \; | \; j \in \{1,\ldots,u_I\}\}$. Since the point $(x,y)$ is not in a section of $\mathscr{C}$, it must be in a sector $C_{I:2j+1}$ of $\mathscr{C}$ for a $j \in \{0,\ldots,u_I\}$. Hence, the real number $y$ belongs to the open interval $U = (\xi_{I:2j}(x), \xi_{I:2(j+1)}(x))$. Since $y$ is in $\partial (S_i)_x$, there exists $z \in U \cap (S_i)_x$ and $z' \in U \cap (S_i)_x^c$. This means that both points $(x,z)$ of $S_i$ and $(x,z')$ of $S_i^c$ belong to $C_{I:2(j+1)}$. This is a contradiction because $\mathscr{C}$ is adapted to $S_i$ (see Lemma \ref{lemma:compl-proj}).

 We now consider a minimal $\mathscr{C} \in \text{CAD}^r(\Fr)$ and prove the other inclusion by contradiction. 
    Suppose that there exists $j \in \{1,\ldots,u_I\}$ such that $\xi_{I:2j}(x)$ is not in $\cup_{i=1}^p\partial (S_i)_x$. Then for  every $i\in\{1,\ldots,p\}$, $\xi_{I:2j}(x)$ lies in the interior of $(S_i)_x$ or in the interior of $(S_i)_x^c$, and there exists $\varepsilon > 0$ such that the open interval $(\xi_{I:2j}(x) - \varepsilon, \xi_{I:2j}(x) + \varepsilon)$ is a subset of $(S_i)_x$ or  $(S_i)_x^c$. 
    This implies that there exists $\eta > 0$ such that the points  $P_\alpha = (x,\xi_{I:2j}(x)+ \alpha \eta), \alpha \in \{-1,0,1\}$ belong simultaneously to $C_{I:2j + \alpha} \cap S_i$ or $C_{I:2j + \alpha} \cap S_i^c$.  Since $\mathscr{C}$ is adapted to $S_i$, the three cells  $C_{I:2j-1}, C_{I:2j},$ $C_{I:2j+1}$ are simultaneously subsets of $S_i$ or of $S_i^c$. Thus, the tuple 
    $\widetilde{\mathscr{C}} =(\mathscr{C}_1,\ldots, \mathscr{C}_{n-1}, \widetilde{\mathscr{C}}_n)$
    with 
    \begin{align*}
        \widetilde{\mathscr{C}}_n = \mathscr{C} _n \setminus \{C_{I:2j-1},C_{I:2j},C_{I:2j+1}\} \cup \{C_{I:2j-1} \cup C_{I:2j} \cup C_{I:2j+1}\}
    \end{align*} is a $\CAD^r$ adapted to $\F$ and strictly smaller than $\mathscr{C}$. This is a contradiction with the minimality of $\mathscr{C}$.
\end{proof}

The previous result asserts that the union of sections of the last level of minimal CADs adapted to $\Fr$ only depends on $\Fr$. We obtain the following result, that will be used in the next section.
\begin{corollary}\label{prop:outil}
    Let $\mathscr{C}$ and $\mathscr{C}'$ be respectively two minimal elements of $\CAD^r(\Fr)$ and $\CAD^{s}(\Fr)$ and consider two cells $C_{I} \in \mathscr{C}_{n-1}, C'_{I'} \in \mathscr{C}'_{n-1}$. If the intersection $C_I \cap C'_{I'}$ is not empty then $u_I = u'_{I'}$ and for all $j \in \{1,\ldots,u_I\}$, the restrictions of the functions $\xi_{I:2j}$ and  $\xi'_{I':2j}$ to $C_I \cap C'_{I'}$ are equal.
\end{corollary}
\begin{proof}
For every $x\in C_I \cap C'_{I'}$, we have 
\[\{\xi_{I:2j}(x) \; | \; j \in \{1,\ldots,u_I\}\}=\cup_{i=1}^p\partial (S_i)_x=\{\xi'_{I:2j}(x) \; | \; j \in \{1,\ldots,u'_{I'}\}\}.\]
Since the sequences $\xi_{I:2j}(x)$ and $\xi'_{I:2j}(x)$ are strictly increasing with $j$, this shows that $u_I=u'_{I'}$ and $\xi_{I:2j}(x)=\xi'_{I:2j}(x)$ for every $j\in\{1,\ldots,u_I\}$.
\end{proof}

%% file: minimum.tex
\section{Minimum CAD{\small s}}\label{sec:minimum}
In this section we investigate the existence of a minimum element in $\text{CAD}^r(\mathcal{F})$. It happens that CADs have properties in low dimensions that are false in higher dimensions (see for instance \cite{lazard2010} or more recently \cite{locatelli-paper}). As we shall see, the properties of the poset $\text{CAD}^r(\mathcal{F})$ depend also heavily on the dimension $n$ of the ambient space.  More precisely, for $n=1$ or $n=2$, we obtain an existence result for every $\mathcal{F}$. However, in dimension $n \geq 3$, we exhibit examples of semi-algebraic sets admitting several distinct minimal adapted CADs and therefore no minimum adapted $\CAD^r$. We describe explicit examples in $\mathbb{R}^3$ admitting several distinct (actually, an infinite set of) minimal CADs of class $C^r, r \in \N \cup \{\infty, \omega\}$ and then use them to produce subsets of $\mathbb{R}^n$ ($n \geq 4$) with no minimum adapted $\CAD^r$. 

\begin{theorem}\label{thrm:existenceMinimum}
For every finite family $\mathcal{F}$ of semi-algebraic sets of $\mathbb{R}$ or of $\mathbb{R}^2$, the poset $(\text{CAD}^r(\mathcal{F}),\preceq)$ admits a minimum element. 
\end{theorem}
\begin{proof}
In view of Proposition \ref{prop:uniqueMin}, it is sufficient to show that if $\mathscr{C}$ and $\mathscr{C}'$ are two minimal elements in $\text{CAD}^r(\mathcal{F})$, then $\mathscr{C} = \mathscr{C}'$. 

Suppose first that the ambient space is $\mathbb{R}$. The projection of any cell of $\mathscr{C}$ or $\mathscr{C}'$ coincides  with $C_{\varepsilon}$, as explained before Proposition \ref{rem:top-caract}. This result directly implies that $\mathscr{C} = \mathscr{C}'$. 

Suppose now that the ambient space is $\mathbb{R}^2$. We first show that $\mathscr{C}_1 = \mathscr{C}'_1$.  Assume for the sake of contradiction that there exists $i \in \{1,\ldots,u_\varepsilon\}$ such that
     $\xi_{2i} \notin \{\xi'_{2j} \; | \; j \in \{1,\ldots,u_\varepsilon'\}\}.$ We will prove that we can merge the corresponding cells of the cylinders over $C_{2i-1}, C_{2i}$ and $C_{2i+1}$ to build $\widetilde{\mathscr{C}}\in \text{CAD}^r(\mathcal{F})$ that is strictly smaller than $\mathscr{C}$, which is a contradiction.  We first show that the semi-algebraic functions of class $C^r$ that define the sections of the cylinders over $C_{2i-1}, C_{2i}$ and $C_{2i+1}$ can be glued together to define cells over the union $C_{2i-1}\cup C_{2i}\cup C_{2i+1}$. Since we have $\xi'_{2}<\cdots<\xi_{2u_\varepsilon'}$,  there exists $j \in \{1,\ldots,u_\varepsilon'+1\}$ such that 
     $\xi'_{2(j-1)} < \xi_{2i} < \xi'_{2j}.$ By definition, $\xi_{2i}$ is in the closure of $C_{2i-1}$, $C_{2i}$ and $C_{2i+1}$, so the intersections $C_{2i-1} \cap C'_{2j-1},
         C_{2i} \cap C'_{2j-1}$ and $C_{2i+1} \cap C'_{2j-1}$ are not empty. Then we can use Corollary \ref{prop:outil} to obtain not only 
     \[u'_{2j-1} = u_{2i-1}= u_{2i}= u_{2i+1},\]
     but also
     \begin{align*}
         \xi'_{2j-1:2k}(x) = \begin{cases}
             \xi_{2i-1 : 2k}(x) &\text{ if } x \in C_{2i-1} \cap C'_{2j-1},\\
             \xi_{2i : 2k}(x) &\text{ if } x \in C_{2i} \cap C'_{2j-1},\\
             \xi_{2i+1 : 2k}(x) &\text{ if } x \in C_{2i+1} \cap C'_{2j-1},
         \end{cases}
     \end{align*}
    for every $k \in \{1,\ldots,u'_{2j-1}\}$. This implies that for any such $k$ the function $\widetilde{\xi}_{2i-1:2k}$, defined on $C_{2i-1} \cup C_{2i} \cup C_{2i+1}$ by
     \[ \widetilde{\xi}_{2i-1:2k}(x) = \begin{cases}
         \xi_{2i-1 :2k}(x) &\text{ if } x \in C_{2i-1},\\
         \xi_{2i : 2k}(x) &\text{ if } x \in C_{2i},\\
         \xi_{2i+1 : 2k}(x) &\text{ if } x \in C_{2i+1},
     \end{cases}\]
     is of class $C^r$, since it coincides with a function of class $C^r$ in the neighbourhood of every point in its domain. It is also semi-algebraic since its graph is the union of the graphs of $\xi_{2i-1:2k}, \xi_{2i:2k}$ and $\xi_{2i+1:2k}$, which are all semi-algebraic by definition.
     Finally, for all $l \in \{1,\ldots,2u'_{2j-1}+1\}$, the three cells $C_{2i-1:l},$ $C_{2i:l}$ and $C_{2i+1:l}$ have respectively a non-empty intersection with the cell $C'_{2j-1:l}$. Using the fact that $\mathscr{C}$ and $\mathscr{C}'$ are adapted to $\mathcal{F}$, we obtain that for every $m\in\{1,\ldots,p\}$, these four cells are simultaneously either a subset of $S_m$ or of $S_m^c$.
     Therefore, the couple $\widetilde{\mathscr{C}} = (\widetilde{\mathscr{C}}_1,\widetilde{\mathscr{C}}_2)$
defined by
     \begin{align*}
         \widetilde{\mathscr{C}}_1 = \mathscr{C}_1 \setminus& \{C_{2i-1},C_{2i},C_{2i+1}\} \cup \{C_{2i-1} \cup C_{2i} \cup C_{2i+1}\},\\
         \widetilde{\mathscr{C}}_2 = \Big(\mathscr{C} _2 \setminus& \big\{C_{2i-1 : 
l},C_{2i:l},C_{2i+1:l} \; | \; l \in \{1,\ldots,2u'_{2j-1}+1\}\big\}\Big) \\
         &\cup \Big\{C_{2i-1:l} \cup C_{2i:l} \cup C_{2i+1:l}\; | \; l \in \{1,\ldots,2u'_{2j-1}\}\Big\}.
     \end{align*}
     is a $\CAD^r$ that is adapted to $\mathcal{F}$ and strictly smaller than $\mathscr{C}$ as requested. This shows that we have the inclusion
     \[\big\{\xi_{2i} \; | \; i \in \{1,\ldots,u_{\varepsilon}\} \big\} \subseteq \big\{\xi'_{2j} \; | \; j \in \{1,\ldots,u_\varepsilon'\}\big\}.\]
     Since the other inclusion follows by symmetry, these finite sets are equal, implying that $u_{\varepsilon} = u_\varepsilon'$ and that $\xi_{2k} = \xi'_{2k}$ for all $k \in \{1,\ldots,u_{\varepsilon}\}.$ In particular, we have $\mathscr{C}_1 = \mathscr{C}_1'$. Using again Corollary \ref{prop:outil} above each cell of $\mathscr{C}_1 = \mathscr{C}_1'$, we obtain that $\mathscr{C}_2 = \mathscr{C}_2'$. 
This means that $\mathscr{C} = \mathscr{C}'$, as announced.
\end{proof}

\begin{remark}\label{rem:top-caract-bis}
    Theorem \ref{thrm:existenceMinimum} and Proposition \ref{rem:top-caract} provide a topological characterization of the minimum of $\CAD^r(\mathcal{F})$ when the ambient space is $\R$. Its set of sections is exactly the union of the boundaries of the elements of $\mathcal{F}$.
\end{remark}

We now present examples of semi-algebraic sets of $\mathbb{R}^3$ admitting several distinct adapted minimal CADs of class $C^r$, hence no minimum adapted $\CAD^r$.  The proofs of Theorem \ref{ex:trousers1} and Corollary \ref{ex:trousersn} presented below are done by hands. As explained in Remark \ref{rem:naive}, the reasoning and techniques will be simplified by using the tools developed in Section \ref{sec:red}.

\begin{definition}
    We define the semi-linear Trousers $\mathbb{T}$ and the analytic Trousers $\mathbb{T}^\omega$ as the semi-algebraic sets given by
     \begin{align*}
        \mathbb{T} &= \left\{(x,y,z) \in \mathbb{R}^3 \;|\; \left((x \leq 0 \lor y \leq 0)\land z = 0\right) \lor \left(x > 0 \land y > 0 \land z = -x/2\right)\right\},\\
        \mathbb{T}^\omega &= \left\{(x,y,z) \in \mathbb{R}^3 \;|\; \left((4 z^4 - 4 z^2 x - y^2 = 0) \land yz < 0\right) \lor \left(y = 0 \land z = 0\right)\right\}.
     \end{align*}
\end{definition}

\begin{figure}[H]
\begin{subfigure}{0.5\textwidth}
    \raisebox{0.3cm}{\includegraphics[scale=0.35]{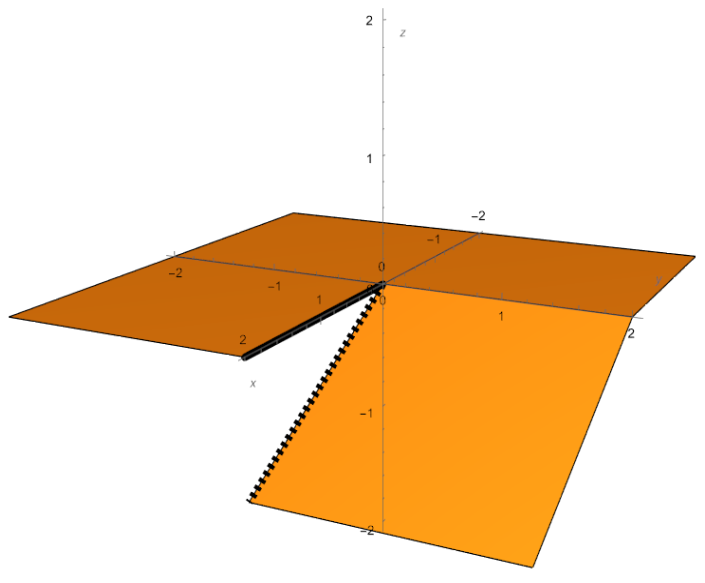}}
\end{subfigure}
\begin{subfigure}{0.5\textwidth}
    \includegraphics[scale=0.33]{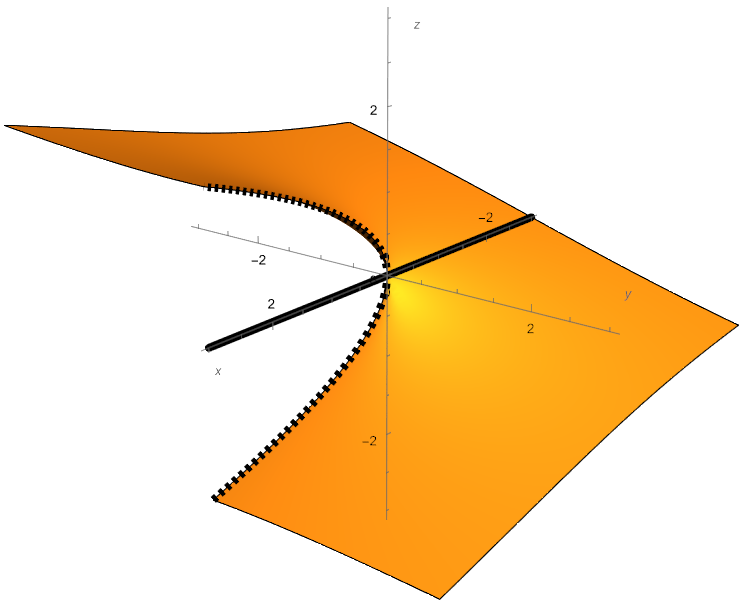}
\end{subfigure}
    \caption{The Trousers $\mathbb{T}$ and $\mathbb{T}^\omega$}
    \label{fig:trousers}
\end{figure}

    Notice that $\mathbb{T}$ is built only with linear polynomials whereas the fourth degree equation involved in the definition of $\mathbb{T}^\omega$ naturally arises when describing the real part of the square root of a complex number $x+ iy$ in Cartesian coordinates.  
    Observe that in Figure~\ref{fig:trousers} the thick half-lines are included in both Trousers, while the dashed curves are not. It happens that the set $\mathbb{T}^\omega$ is not an analytic embedded submanifold of $\R^3$, but it is the union of two. 


\begin{figure}[H]
        \center
        \includegraphics[scale=0.25]{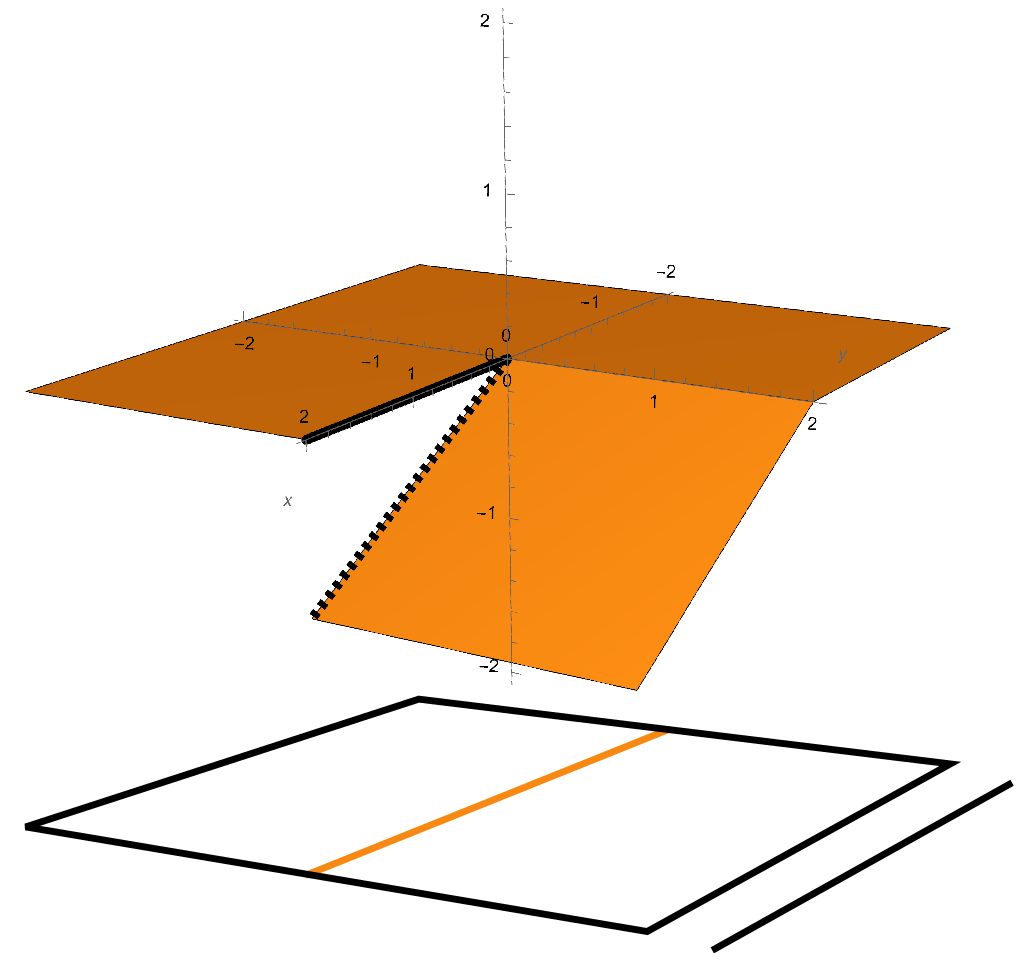} \; \;
        \includegraphics[scale=0.25]{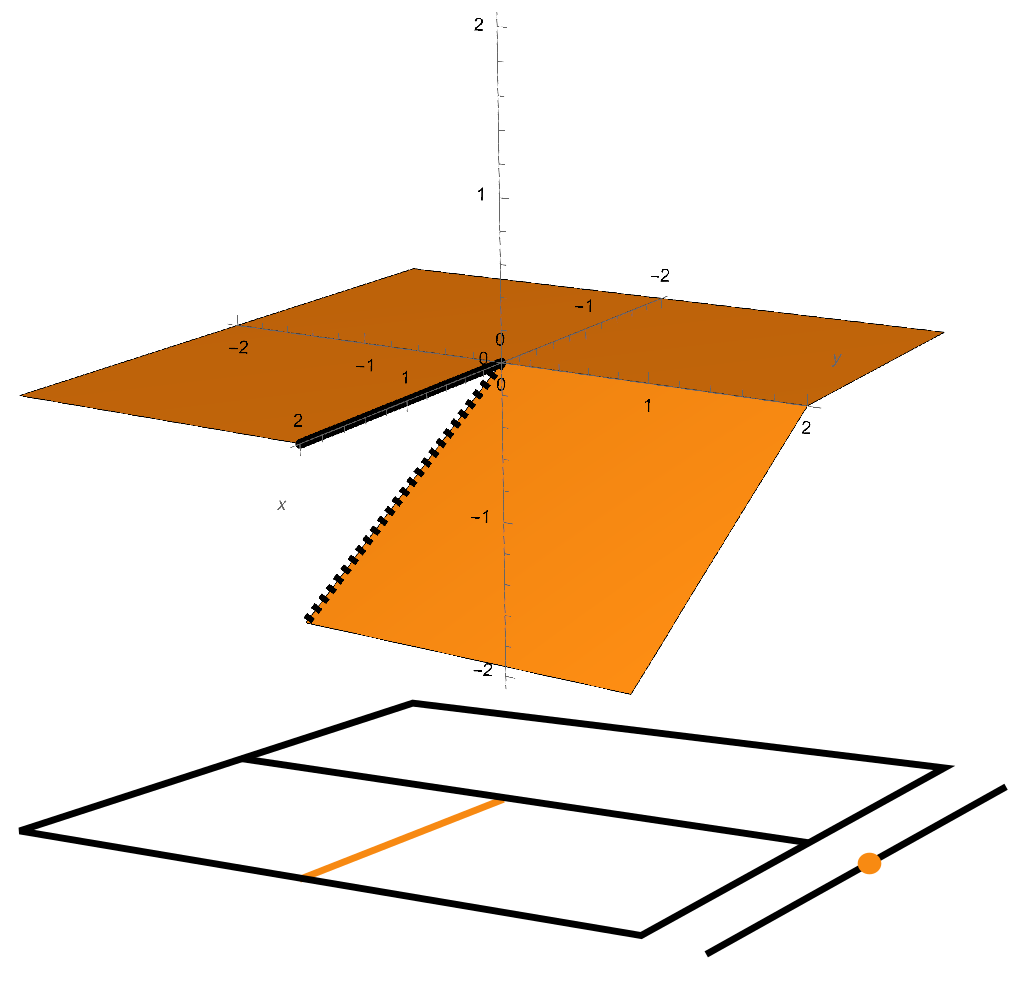}
        \caption{The CADs $\mathscr{C}$ and $\mathscr{C}'$}
        \label{fig:trousersCADs}
    \end{figure}

\begin{theorem}\label{ex:trousers1}
    For $r \in \N \cup \{\infty, \omega\}$, the posets $(\CAD^0(\mathbb{T}), \preceq)$ and $(\CAD^r(\mathbb{T}^\omega), \preceq)$ admit no minimum element.
\end{theorem}
\begin{proof}
We describe two distinct minimal elements of $\CAD^0(\mathbb{T})$ as depicted in Figure \ref{fig:trousersCADs}, and conclude by Proposition \ref{prop:uniqueMin}. To this end, we consider the function $f : \R^2 \to \R$ defined by 
\begin{equation}\label{eqn:f}
    f(x,y) = -\frac{x}{2}\chi(x,y),
\end{equation} where $\chi$ is the indicator function of the set $\{(x,y) \in \mathbb{R}^2 \; | \; x > 0 \land y > 0\}$.
For the first one, the partition $\mathscr{C}_1$ contains the unique cell
$C_1 = \mathbb{R}$ while $\mathscr{C}_2$ consists in the three cells $C_{11} = C_1 \times (-\infty, 0), C_{12} = C_1 \times \{0\}, C_{13} = C_1 \times (0,+\infty).$ The partition $\mathscr{C}_3$ consists in the nine cells obtained by slicing exactly once the cylinder above each of these cells with appropriate restrictions (which are continuous) 
    of the function $f$.
For the second one, the partition $\mathscr{C}'_1$ contains the three cells
$C'_1 = (-\infty,0)$, $C'_2 = \{0\}$, $C'_3 = (0,+\infty)$ while $\mathscr{C}'_2$ consists in the cells $C'_{11} = C'_1 \times \mathbb{R}, C'_{21} = C'_2 \times \mathbb{R}, C'_{31} = C'_3 \times (-\infty, 0), C'_{32} = C'_3 \times \{0\}, C'_{33} = C'_3 \times (0,+\infty).$ The partition $\mathscr{C}'_3$ is built with the fifteen cells in the same way as $\Cr_3$: we slice exactly once the cylinder above each of the  cells of $\Cr_2'$ with appropriate restrictions (which are continuous) 
    of the function $f$.
     These CADs $\mathscr{C}$ and $\mathscr{C}'$ of class $C^0$, which are obviously distinct, are adapted to $\mathbb{T}$ because
     \begin{align*}
         \mathbb{T} = C_{112} \cup C_{122} \cup C_{132} = C'_{112} \cup C'_{212} \cup C'_{312}\cup C'_{322} \cup C'_{332}.
     \end{align*}
     A careful but direct inspection then shows that these CADs are indeed minimal (see also Remark \ref{rem:trousersHard} below). 

    For every $r\in \N\cup \{+\infty, \omega\}$, we proceed as above and construct two distinct minimal elements of $(\CAD^r(\mathbb{T}^\omega), \preceq)$. We consider the same CADs $\Cr$ and $\Cr'$ as before, but instead of $f$, we use the function $g : \R^2 \to \R$ defined by 
    \begin{equation}\label{eqn:g}
        g(x,y) = - \sign(y)  \sqrt{\frac{x + \sqrt{x^2 + y^2}}{2}},
    \end{equation}
    where $\sign(y)$ is defined by $1$ (resp. $0, -1$)  if $y>0$ (resp. $y=0, y<0$). It is readily seen that the restriction to every cell of $\Cr_2$ and $\Cr_2'$ (except for $C_{11}'$) of the function $g$ is analytic, hence of class $C^r$. For the cell $C_{11}'$, 
    we use the analytic change of coordinates (polar coordinates) given by $\varphi(\rho, \theta) = (\rho\cos(\theta), \rho\sin(\theta)), \rho>0, \theta \in (0,2\pi)$ and  observe that the function $g\circ \varphi$ satisfies $g\circ \varphi(\rho,\theta) =\sqrt{\rho} \cos\left(\frac{\theta}{2}\right)$, which is analytic. 
    \end{proof}

    \begin{remark}\label{rem:trousersHard}
        In order to show the minimality of $\mathscr{C}$ and $\mathscr{C}'$ in the previous proof by considering all the partitions that are strictly smaller than $\mathscr{C}$ and $\mathscr{C}'$, one is led to investigate the sets $\text{SSP}(\mathscr{C})$ and $\text{SSP}(\mathscr{C}')$ which contain respectively $B_9 - 1 =  21\,146$ and $B_{15} - 1 =  1\, 382\, 958\,
        544$ elements, as discussed in Remark \ref{rem:naive}. Taking into account the properties of CADs (e.g. the topology of cells and the cylindrical character of the partition) drastically reduces the number of partitions to be taken into consideration. Hence, the brute force approach is doomed to be inefficient.
        This observation leads us to study in the next section the notion of CAD reduction. This concept will enable us to analyse the order on $\CAD^r(\Fr)$, and for a given $\CAD^r$ $\mathscr{C}$, quickly remove many candidates from SSP($\mathscr{C}$) that will never give rise to a $\CAD^r$. We will then easily justify the minimality of the CADs given in the proof of Theorem \ref{ex:trousers1} (see Example \ref{ex:trousers2}).
    \end{remark}

    We proved the previous proposition by showing that the Trousers $\mathbb{T}$ (and $\mathbb{T}^\omega$) has at least two distinct minimal CADs. 
    We now strengthen this result. 

    \begin{proposition}\label{prop:infinity}
        For $r \in \N \cup \{\infty, \omega\}$, the posets $(\CAD^0(\mathbb{T}), \preceq)$ and $(\CAD^r(\mathbb{T}^\omega), \preceq)$ admit infinitely many pairwise distinct minimal elements.
    \end{proposition}
    \begin{proof}
        For every real algebraic number $t$ such that $t \leq 0$, we construct a minimal element $\Dr^{t}$ of $\CAD^0(\mathbb{T})$. Intuitively, the construction follows the same approach as the one of the minimal $\CAD^0$ $\Cr'$ adapted to $\mathbb{T}$ defined in the proof of Theorem \ref{ex:trousers1}, except that the section at the first level is $\{t\}$ instead of $\{0\}$. In particular, the CAD $\Dr^0$ coincides with $\Cr'$.
        
        The partition $(\Dr^t)_1$ contains the unique section $\{t\}$ and $(\Dr^t)_2$ has a unique section defined by $(t, + \infty) \times \{0\}$. 
        The partition $(\Dr^t)_3$ is obtained as in the proof of Theorem \ref{ex:trousers1} by slicing the cylinder above every cell of $(\Dr^t)_2$ with the corresponding restriction of the function $f$ defined in Equation \eqref{eqn:f}. 
        By following the same reasoning as for the CAD $\Cr'$ (see Theorem \ref{ex:trousers1} and the subsequent remark), we obtain that $\Dr^t$ is a minimal element of $\CAD^0(\mathbb{T})$. The result follows since, for any two distinct negative real algebraic number $s$ and $t$, we have $(\Dr^s)_1 \neq (\Dr^t)_1$, and hence $\Dr^s \neq \Dr^t$.
        
        We proceed similarly to construct an infinite collection of minimal elements of $\CAD^r(\mathbb{T}^\omega)$, by using the function $g$ (see Equation \eqref{eqn:g}) instead of $f$.
    \end{proof}
    \begin{remark}
        The description of minimal elements of $\CAD^0(\mathbb{T})$ may be generalized by considering, for every real algebraic number $t \leq 0$, the set $\mathcal{A}_t$ of continuous semi-algebraic functions $a : (t, +\infty) \to \R$ satisfying $a(x) = 0$ if $x \geq 0$. We construct a minimal element $\Er^{t,a}$ in a similar manner as $\Dr^t$ (see the proof of Proposition \ref{prop:infinity}). We define $(\Er^{t,a})_1$ as $(\Dr^t)_1$, and $(\Er^{t,a})_2$ as the $\CAD^0$ of $\R^2$ with a unique section defined by the graph of $a$. Finally, $(\Er^{t,a})_3$ is obtained similarly to $(\Dr^t)_3$, by slicing the cylinder above every cell of $(\Er^{t,a})_2$ using the corresponding restriction of $f$. Note that if $z_t$ is the constant function $0$ on $(t, +\infty)$, then $\Er^{t,z_t} = \Dr^t$.
        This method extends to any regularity $r \in \N \cup \{\infty\}$ for $\mathbb{T}^\omega$, but does not allow to describe additional minimal elements of $\CAD^\omega(\mathbb{T}^\omega)$, since for all $t \leq 0$, there exists only one analytic function $a : (t, +\infty) \to \R$ such that $a(x) = 0$ if $x \geq 0$, namely the constant function $z_t$.
    \end{remark}

    We now extend Theorem \ref{ex:trousers1} in higher dimension. For $n \geq 4$, we denote by $\mathbb{T}_n$ and $\mathbb{T}^\omega_n$ the semi-algebraic sets $\mathbb{T} \times \R^{n-3}$ and $\mathbb{T}^\omega \times \R^{n-3}$.

    \begin{corollary}\label{ex:trousersn}
            For $n \geq 4, r \in \N \cup \{\infty, \omega\}$, the posets $(\CAD^0(\mathbb{T}_n), \preceq)$ and $(\CAD^r(\mathbb{T}^\omega_n), \preceq)$ admit no minimum element.
    \end{corollary}
    \begin{proof}
    We focus on the case of $\mathbb{T}_n$ since the others are similar. By the very definition of $\mathbb{T}_n$, we have $(x_1,x_2, x_3) \in \mathbb{T}$ if and only if for all $x_4,\ldots,x_n \in \mathbb{R}$, $(x_1,\ldots,x_n) \in \mathbb{T}_n$. It follows that if $\mathcal{D} = (\mathcal{D}_1, \mathcal{D}_2, \mathcal{D}_3)$ is a $\CAD^0$ adapted to $\mathbb{T}$, then the tuple $\widetilde{\mathcal{D}} = ( \mathcal{D}_1, \mathcal{D}_2, \mathcal{D}_3, \mathcal{D}_4, \ldots, \mathcal{D}_n)$, where
     \[\mathcal{D}_k = \big\{D \times \mathbb{R} \; | \; D \in \mathcal{D}_{k-1}\big\}\]
     for $k \geq 4$, is a $\CAD^0$ adapted to $\mathbb{T}_n$. Considering again the CADs $\mathscr{C}$ and $\mathscr{C}'$ adapted to $\mathbb{T}$ described in the proof of Theorem \ref{ex:trousers1}, a careful inspection shows that the corresponding CADs $\widetilde{\mathscr{C}}$ and $\widetilde{\mathscr{C}}'$ are two distinct minimal elements of $\CAD^0(\mathbb{T}_n)$.  
\end{proof}

We now exhibit more examples of semi-algebraic sets that do not admit minimum CADs of class $C^r$. Each of these is interesting in its own right and their consideration shows that the question of singling out a topological property of the semi-algebraic set $S$ that would guarantee that the poset $\CAD(S)$ admits a minimum element is not an elementary one and would deserve a careful study. These constructions can be readily extended to higher dimensions.

\begin{example}\label{ex:counterclosed}
     Neither of the closed semi-algebraic sets
     \begin{align*}
         \mathcal{B} = &[-1,1] \times [-1,1] \times [0,1] \\
         &\cup \big\{(x,y,z) \in \mathbb{R}^3 \; | \; y = 0, 0 \leq x \leq 1, 1 \leq z \leq x+1\big\},\\
         \mathcal{U} = &\big\{(x,y,z) \in \mathbb{R}^3 | \; (x \leq 0 \lor y \leq 0)  \land z \leq 0\big\}\\
             & \cup \big\{(x,y,z) \in \mathbb{R}^3 |\; x \geq 0 \land y \geq 0 \land 
x+yz \leq 0 \land z \leq 0\big\}
     \end{align*}
     admits a minimum $\CAD^0$. Observe that $\mathcal{B}$ is bounded while $\mathcal{U}$ is unbounded. We can indeed describe two distinct minimal CADs adapted to $\mathcal{U}$ using the same  $\mathscr{C}$ and $\mathscr{C}'$ defined for the Trousers in the proof of Theorem \ref{ex:trousers1}, but replacing the function $f$ by $h$ defined by $h(x,y)=-\frac{x}{y}\chi(x,y)$. The construction of two minimal CADs adapted to $\mathcal{B}$ follows the same lines with suitable adaptations.  It is therefore omitted. 
     
     \begin{figure}[H]
         \center
         \includegraphics[scale=0.37]{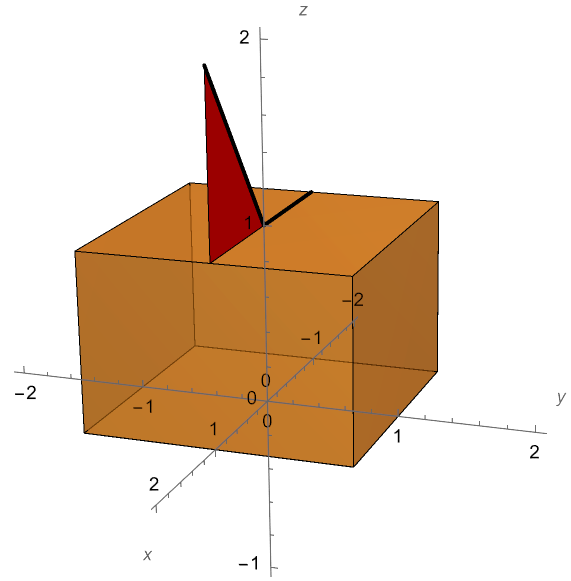}~
         \includegraphics[scale=0.34]{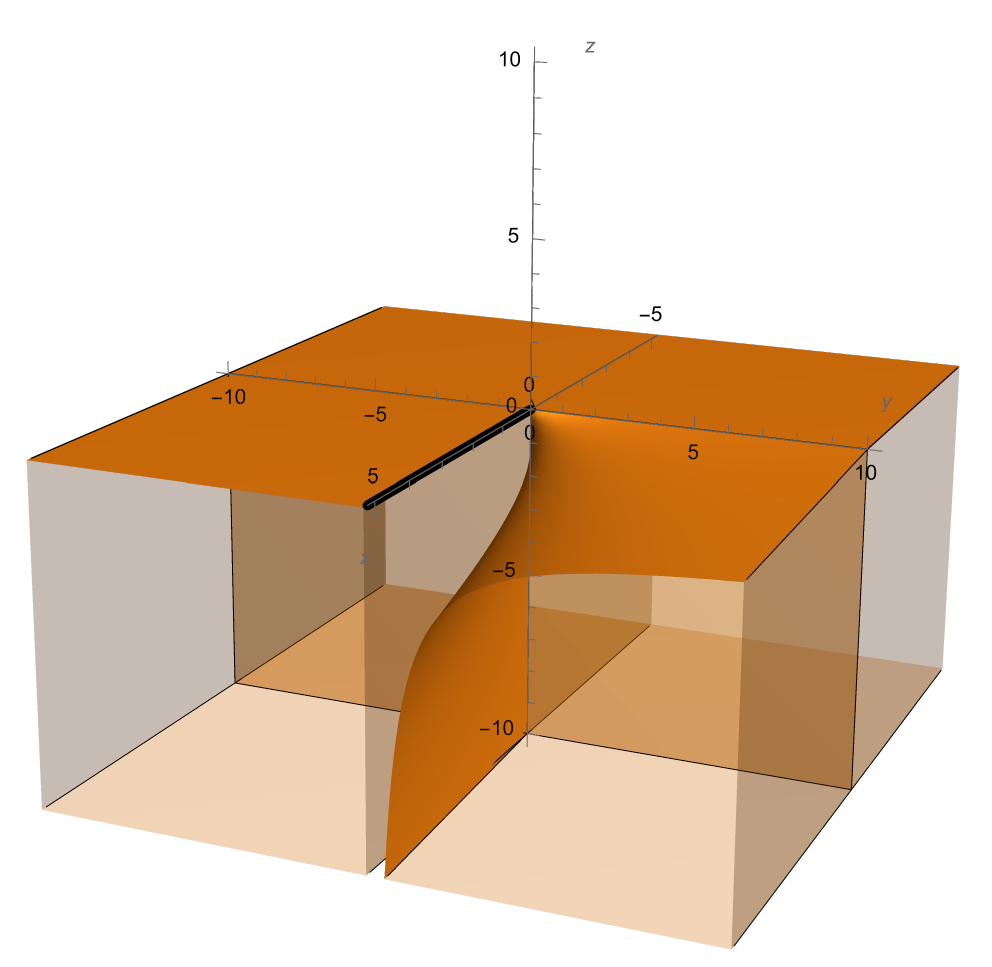}
         \caption{Semi-algebraic sets with two distinct minimal CADs: $\mathcal{B}$ (left) and $\mathcal{U}$ (right)}
     \end{figure}
     
Observe also that Lemma \ref{lemma:compl-proj} guarantees that neither of the open semi-algebraic sets defined as the complement of the closed sets $\mathcal{B}$ and $\mathcal{U}$ considered in Example \ref{ex:counterclosed} admits a minimum $\CAD^0$. 
     
\end{example}

\begin{example}
For every $r \in \N \cup \{\infty, \omega\}$, the semi-algebraic sets defined by
\begin{align*}
    \mathbb{A} &= \left\{(x,y,z) \in \R^3 \; | \; \left(x^2 + 4yz = 0 \land \left(x \neq 0 \lor y \neq 0\right)\right) \lor \left(y = z = 0\right)\right\},\\
    \Pi &= \left\{(x,y,z) \in \R^3 \; | \; (z = 0 \land y \neq 0) \lor (z = -x \land y = 0)\right\}
\end{align*} 
 and depicted in Figure \ref{fig:analytic} do not admit a minimum $\CAD^r$. 
    We indeed build two distinct minimal elements $\Cr$ and $\Cr'$ of  $\CAD^r(\mathbb{A})$ using the same partitions $\Cr_1, \Cr_2$ and $\Cr_1'$ defined in the proof of Theorem \ref{ex:trousers1}. The $\CAD^r$ $\Cr_2'$ is obtained by slicing the cylinder above $C'_1$ and $C'_3$ by the unique section given by constant function $0$.
    Finally, we define the corresponding decompositions of $\R^3$ by slicing exactly once the cylinder above each of these cells with appropriate restrictions (which are of class $C^\omega$) 
    of the function $\eta$, that vanishes when $y = 0$, and defined by $\eta(x,y) = -\frac{x^2}{4y}$ otherwise. To show that these two distinct CADs are indeed minimal elements of $\CAD^r(\mathbb{A})$, we proceed as in Example \ref{ex:trousers2}. Similarly, we construct two distinct minimal elements of $\CAD^r(\Pi)$ following the same approach, but replacing $\eta$ with $\gamma$ defined by $\gamma(x,0) = -x$ and $\gamma(x,y)=0$ if $y\neq 0$. 
\begin{figure}[H]
    \center
    \includegraphics[scale=0.35]{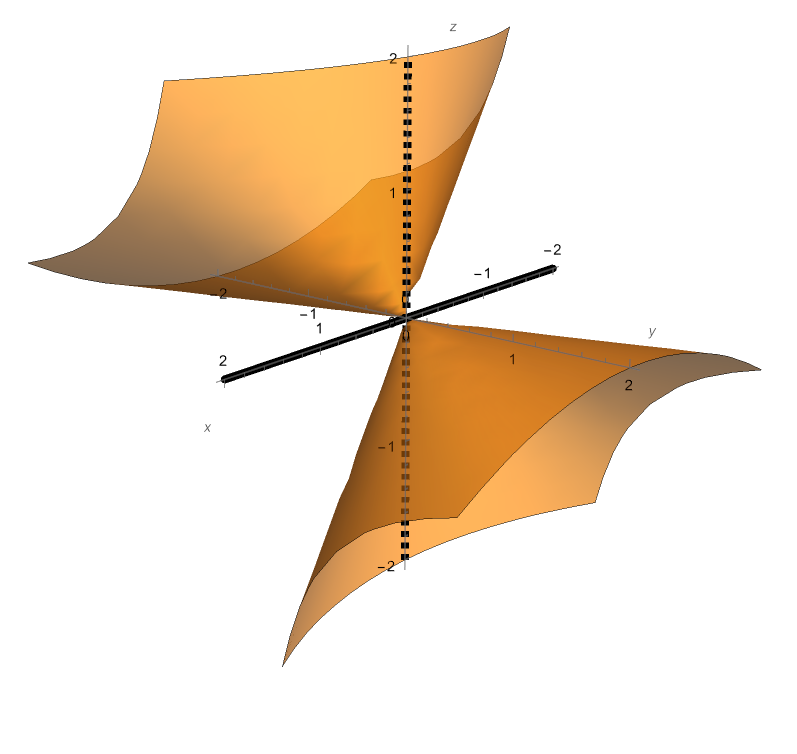}~\includegraphics[scale=0.35]{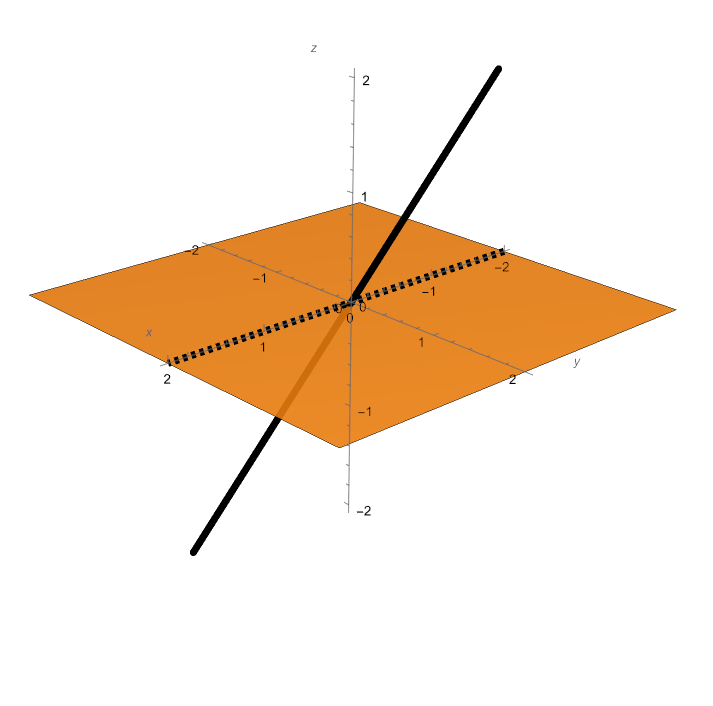}
    \caption{The semi-algebraic sets $\mathbb{A}$ (left) and $\Pi$ (right)}
    \label{fig:analytic}
\end{figure}
\end{example}

%% file: reductions.tex
\section{CAD reductions}\label{sec:red}

The determination of CADs that are adapted to a a family $\Fr$ and smaller than a given CAD $\mathscr{C}$ by direct inspection is certainly inefficient and already cumbersome even for low dimensional examples (see Section \ref{sec:minimum} and more precisely Remark \ref{rem:trousersHard}). The notion of CAD reduction that is developed in this section offers a better grasp on the partially ordered set ($\text{CAD}^r(\Fr), \preceq$). From a practical point of view, it enables us to introduce an algorithm for simplifying the computation of minimal CADs of class $C^r$. We apply this algorithm to the CADs introduced in the proof of Theorem \ref{ex:trousers1}, and provide a short proof of their minimality. From a theoretical standpoint, we show that the relation defined by $\CAD^r(\Fr)$ reductions is actually the transitive reduction (in the sense of \cite{TR1972}) of the order relation $\preceq$ on $\CAD^r(\Fr)$. We finally obtain a characterization of those posets $(\CAD^r(\Fr), \preceq)$ that admit a minimum in terms of the confluence of the associated abstract reduction system $(\CAD^r(\Fr), \leftarrow)$. 

\subsection{Tree structure of a CAD}
The tree associated with a given CAD $\mathscr{C}$ can be seen as a simplified version of $\mathscr{C}$, which only encodes its combinatorial structure. Roughly speaking, it is defined by considering only the cell indices of the considered CAD and a function that encodes the inclusion of the cells in the sets $S_i \in \Fr$ under consideration. 
We first introduce the general notion of a CAD tree.

\begin{definition}\label{def:CADTREE}
    A CAD tree of depth $n$ and width $p$ is a pair $\mathcal{T} = (T,L)$ where 
        \begin{itemize}
            \item $T$ is a labelled rooted odd-ary tree whose nodes are labelled by tuples. The root is labelled by the empty tuple $\varepsilon$ and if $I$ is the label of a node, then either $I$ is an $n$-tuple and the node is a leaf, or there exists $u_I \in \mathbb{N}$ such that the node has exactly $2u_I + 1$ sons labelled by $I : j$ with  $j \in \{1,\ldots,2u_I + 1\}$. As we continue, we identify a node with its label, thus seeing $T$ as a subset of $\bigcup_{k=0}^n(\mathbb{N}^*)^k$ endowed with the prefix order relation.
            \item $L$ is a map defined recursively on $T$ from the leaves to the root by
                \[L(I) \begin{cases}
                    \in \{0,1\}^p &\text{ if $I$ is a leaf of $T$,}\\
                    = \big(L(I:1), \ldots, L(I:2u_I+1)\big) &\text{ otherwise.}
                \end{cases} \]   
        \end{itemize}    
\end{definition}

    The redundancy in the definition above is useful for writing \eqref{eqn:L} below easily. However, it is clear that the function $L$ is completely defined by its values on the leaves of~$T$.



\begin{definition}\label{def:treeCAD}
    For every $\mathscr{C} \in \text{CAD}^r(\Fr)$, the CAD tree associated with $(\mathscr{C}, \Fr)$ is the CAD tree of depth $n$ and width $p$ where the nodes of $T$ are the indices of the cells of $\mathscr{C}$ and where the function $L$ is defined on the leaves of $T$ by the $p$-tuple $L(I)$ whose $i^\text{th}$ component is equal to $1$ if $C_I \subseteq S_i$ and by $0$ otherwise. We denote it by Tree$(\mathscr{C} ,\Fr)$ or simply Tree$(\mathscr{C})$ if there is no ambiguity.
\end{definition}

\begin{example}\label{ex:diskMotiv} 
    Let $S$ be the closed disk centred at the origin and of radius one. 
    Consider the CAD $\mathscr{C}$ adapted to $S$ where the five cells of $\mathscr{C}_1$ are defined by $C_1 = (-\infty,-1)$, $C_2 = \{-1\},$ $C_3 = (-1,1)$, $C_4 = \{1\}$, $C_5 = (1,+\infty)$ and where the cylinders above these cells are split by exactly the two functions defined on $C_2$ and $C_4$ by $0$ and by the two functions defined on $C_3$ by $-\sqrt{1-x^2}$ and $\sqrt{1-x^2}$.
    We also consider another CAD $\mathscr{C}'$ adapted to $S$. It is obtained by splitting the cell $C_3$ into three cells $ (-1,0), \{0\}, (0,1)$. The functions used to cut cylinders of this new CAD are the same as those for $\mathscr{C}$, subject to appropriate restrictions. 
    Figure \ref{fig:disk} represents these CADs and their associated CAD trees. 
    A leaf $I$ such that $L(I)$ is $0$ (resp. $1$) is represented in red (resp. green). Moreover, we shorten the labelling of the nodes in an obvious manner.
    \begin{figure}[H]
            \includegraphics[width=0.48\linewidth]{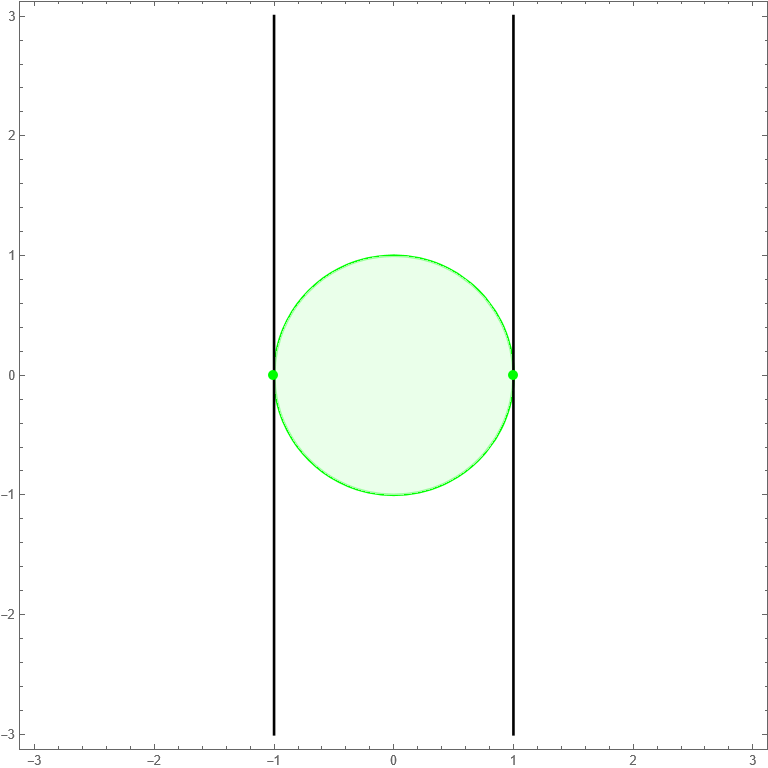}
            \includegraphics[width=0.48\linewidth]{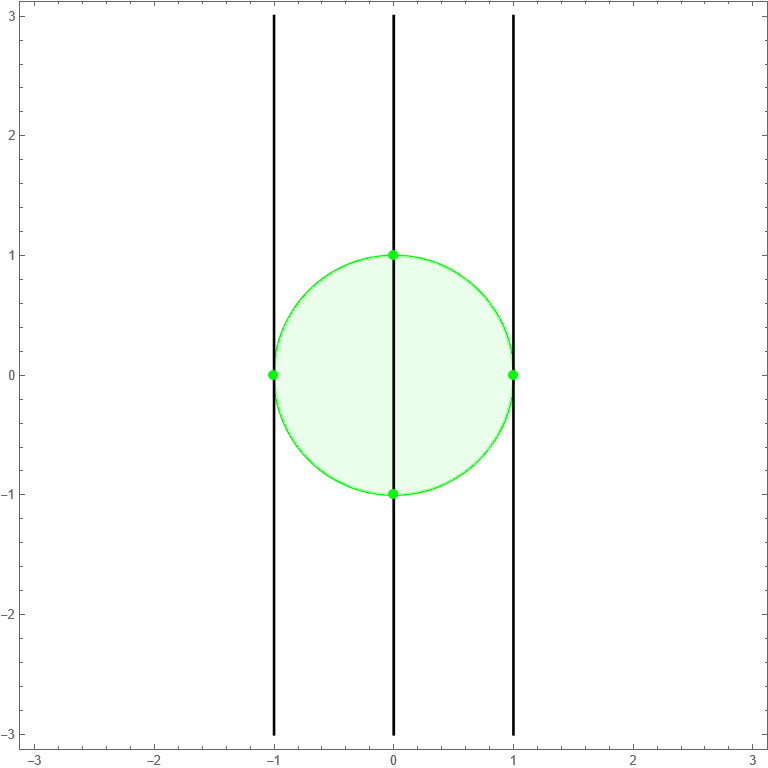}
              \includegraphics[width=\linewidth]{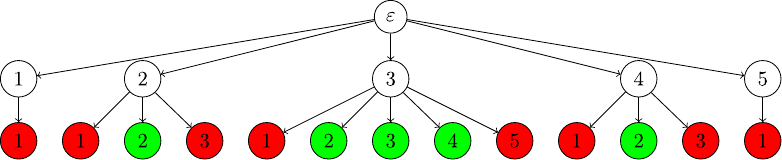}
        \includegraphics[width=\linewidth]{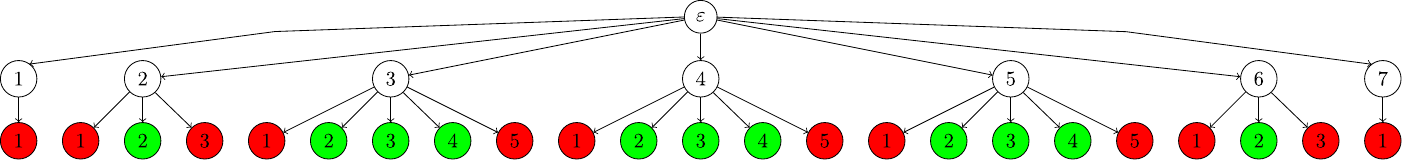}
            \caption{The CADs $\mathscr{C}$ and $\mathscr{C}'$ and their associated CAD trees}
            \label{fig:disk}
    \end{figure}

\end{example}

\subsection{Reductions}
The concepts of reductions, which we now introduce, may already be seen in Example \ref{ex:diskMotiv}. Indeed, the CAD $\mathscr{C}$ is obtained by merging corresponding cells of $\mathscr{C}'$ above the third, fourth and fifth cylinders. Furthermore, the possibility to merge these cells of $\mathscr{C}'$ implies the following on the associated CAD trees:
\begin{itemize}
    \item the subtree with prefix 3 of Tree$(\mathscr{C})$ is identical to the subtrees with prefixes 3,4 and 5 of Tree$(\mathscr{C}')$;
    \item the subtrees with prefixes 4 and 5 of Tree$(\mathscr{C})$ are respectively identical to the subtrees with prefixes 6 and 7 of Tree$(\mathscr{C}')$. 
\end{itemize}
In this example, we will say that there is a reduction from $\text{Tree}(\mathscr{C}')$ to $\text{Tree}(\mathscr{C})$. It is obtained by removing subtrees with prefixes 4 and 5 and properly relabelling the other nodes (by subtracting $2$ to the 1-prefixes of subtrees with prefixes 6 and 7). In order to perform these operations in the general case, we make use of the $k^\text{th}$ prefix map
    \[p_{k} \colon \bigcup_{l=0}^{+\infty}(\mathbb{N}^*)^l \to \bigcup_{l=0}^{+\infty}(\mathbb{N}^*)^l \colon (i_1,\ldots,i_l) \mapsto \begin{cases}
        (i_1,\ldots,i_l)&\text{ if $l < k$},\\
        (i_1,\ldots,i_k)&\text{ if $l\geq k$}.
       \end{cases}
    \]
When we merge a section with (even) index $A$ with two surrounding sectors in a level of a CAD, we have to merge the subtrees corresponding to these cells and relabel the whole tree accordingly. This is the purpose of the functions we now introduce.

\begin{definition}\label{def:auxpsi} 
    For every even tuple $A \in (\mathbb{N}^*)^k$ ($k \geq 1$),  we define 
    {\small\begin{align*}
        \psi_A \colon \bigcup_{l=0}^{+\infty}(\mathbb{N}^*)^l \to \bigcup_{l=0}^{+\infty}(\mathbb{N}^*)^l \colon I \mapsto \begin{cases}
            I - e_k \text{ if } p_k(I) = A,\\
            I - 2 e_k \text{ if } \exists m \in \mathbb{N}^* :  p_k(I) = A+me_k,\\
            I \text{ otherwise},
        \end{cases}
    \end{align*}}
    where $e_k$ is the $k^\text{th}$ unit vector of $\mathbb{R}^l$ for $l \geq k$.
\end{definition}

\begin{definition}\label{def:CADtreeRed}
    Consider a CAD tree $\mathcal{T} = (T,L)$ and an even node $A \in T \cap (\mathbb{N}^*)^k$ ($k \geq 1$). We say that $\psi_A$ induces a reduction rule on $\mathcal{T}$ if we have
    \begin{equation}\label{eqn:L}
             L(A - e_k) = L(A) = L(A + e_k).
         \end{equation}
 Then the induced reduction rule is denoted by $\Psi_{A}$ and the reduced CAD tree is given by $\mathcal{T}' =  (\psi_{A}(T), L')$ where $L'$ is defined on the leaves of $\psi_A(T)$ by $L'(\psi_A(I))=L(I)$ for every leaf $I$ of $T$. We write\footnote{Here we make use of the terminology on Abstract Reduction Systems from \cite{TRaAT}.} $\mathcal{T} \to \mathcal{T'}$ or $\mathcal{T}' \leftarrow \mathcal{T}$.
 \end{definition}

 The definition of $\psi_A$ and Condition \eqref{eqn:L} ensure that $\mathcal{T}'$ is a well-defined CAD tree. The existence of a reduction rule $\Psi_{A}$ from Tree$(\mathscr{C})$ characterizes the fact that the cylinders above $C_{A - e_k}, C_A$ and $C_{A + e_k}$ of $\mathscr{C}$ contain the same number of cells and the corresponding ones are all in $S_i$ or all in $S_i^c$, for every $i\leq p$. We will say that $\Psi_A$ induces a $\CAD^r$ reduction if we can ensure that the unions of cells prescribed by $\Psi_A$ give rise to a $\CAD^r$.

\begin{definition}\label{def:CADRed}
    Let $\mathscr{C}$ be in $\text{CAD}^r(\Fr)$.
    A CAD tree reduction rule $\Psi_{A}$ from $\text{Tree}(\mathscr{C})$ to $\mathcal{T}'$ lifts to a $\text{CAD}^r(\Fr)$ reduction rule $\Phi_A$ defined on $\mathscr{C}$ if 
    \[\mathscr{C}' = \left\{\bigcup_{I \in \text{Tree}(\mathscr{C}) \; : \; \psi_A(I) = I'} C_{I} \; \Big| \; I' \text{ leaf of } T'\right\}\]
    is a CAD  of class $C^r$. In this case, the reduced $\CAD^r$ is given by $\mathscr{C}'$ and we write $\mathscr{C} \to \mathscr{C}'$ or $\mathscr{C}' \leftarrow \mathscr{C}$.
\end{definition}

The existence of the CAD tree reduction rule $\Psi_{A}$ is a necessary condition for the existence of the corresponding $\CAD^r(\Fr)$ reduction rule $\Phi_{A}$. It provides an algorithmically-friendly way to generate the candidates for $\text{CAD}^r(\Fr)$ reductions, depending only on the combinatorial structure of the CAD. When this necessary condition is satisfied, the existence of $\Phi_A$ depends on the geometry of the CAD, in particular, but not only, on the adjacency of the cells.

\begin{figure}[H]
        \center 
        \includegraphics[scale=0.22]{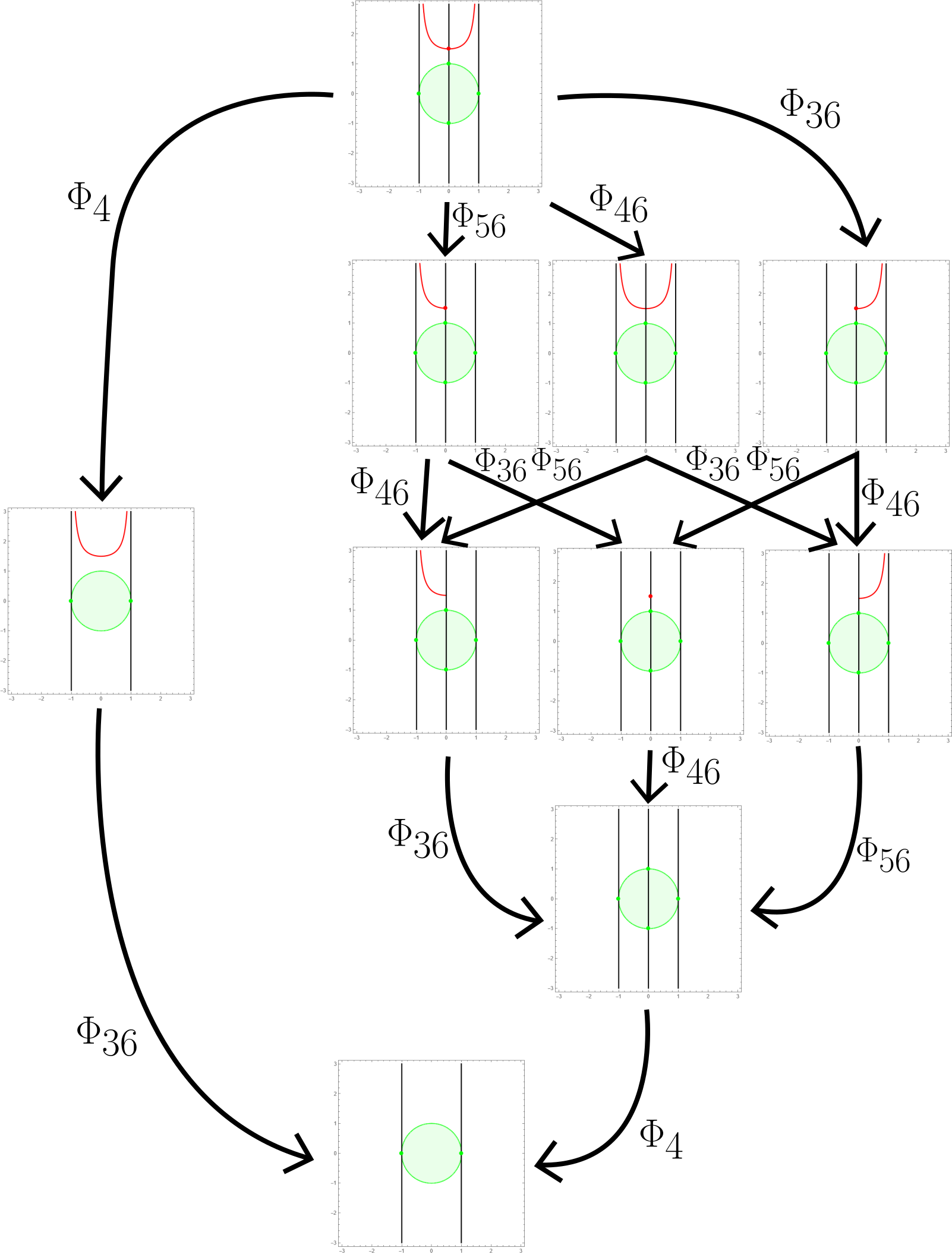}
        \caption{Hasse diagram of elements of $\text{CAD}^r(S)$ smaller than or equal to $\mathscr{C}''$}
        \label{fig:hasse}
    \end{figure}

\begin{example}
    In Example \ref{ex:diskMotiv}, the reduction rule $\Psi_4$ from Tree$(\mathscr{C}')$ to Tree$(\mathscr{C})$ exists and lifts to a reduction rule $\Phi_4$ from $\mathscr{C}'$ to $\mathscr{C}$. The reduction rule $\Phi_4$ consists in merging together the corresponding cells from the third, fourth and fifth cylinders of $\mathscr{C}'$, giving the CAD $\mathscr{C}$. 
    An example of a CAD tree reduction rule that does not lift to CAD reduction rule is studied in Example \ref{ex:trousers2}.
\end{example}
To relate the reduction relation $\leftarrow$ with the order $\preceq$, we need the following lemma.
\begin{lemma}\label{lemma:tradOrder}
    Let $\Cr, \Cr' \in \CAD^r(\Fr).$ 
    \begin{enumerate}
        \item If $\Cr'\preceq \Cr$, then $\Cr'_k \preceq \Cr_k$ for all $k \in \{1, \ldots, n\}$.
        \item If $\Cr'\prec \Cr$, then there exists $l \in \{1, \ldots, n\}$ such that  $
        \mathscr{C}'_k = \mathscr{C}_k$ if $k < l$ and $\mathscr{C}_{k}' \prec \mathscr{C}_k$ if $k \geq l$.  
    \end{enumerate} 
\end{lemma}
\begin{proof}
    For every cell $C'_K\in \Cr'_k$ there exists $C'\in \Cr'_n$ whose projection on $\R^k$ is $C'_K$. Since $\Cr' \preceq \Cr$, $C'$ is a union of cells of $\Cr$ and so $C_K'$ is the union of the projections of these cells onto $\R^k$, which belong to $\Cr_k$. Thus we have $\Cr'_k \preceq \Cr_k$. If $\Cr' \prec \Cr$ (i.e. $\Cr'_n \prec \Cr_n$), we have $\Cr'_k \preceq \Cr_k$ and there exists a smallest $l$ such that $\Cr'_l \prec \Cr_l$. Then by definition, we have $\mathscr{C}'_k = \mathscr{C}_k$ if $k < l$, and for $k\geq l$ we must have $\mathscr{C}_{k}' \prec \mathscr{C}_k$, because otherwise we would have $\mathscr{C}_{k}' = \mathscr{C}_k$, and by projection $\mathscr{C}_{l}' = \mathscr{C}_l$, which is a contradiction.
\end{proof}
The following theorem enables us to study the poset $\text{CAD}^r(\Fr)$ via CAD reductions.
We denote by $\stackrel{*}{\leftarrow}$ the reflexive and transitive closure of the relation $\leftarrow$ on $\text{CAD}^r(\Fr)$.
\begin{theorem}\label{prop:lien-red-ordre}
    Let $\mathscr{C}, \mathscr{C} ' \in \text{CAD}^r(\Fr)$. We have $\mathscr{C}' \stackrel{*}{\leftarrow} \mathscr{C}$ if and only if $\mathscr{C}' \preceq \mathscr{C}.$
\end{theorem}
\begin{proof} By definition of CAD reductions, if $\mathscr{C}' \stackrel{*}{\leftarrow} \mathscr{C}$, then all the cells of $\mathscr{C}'$ are unions of cells of $\mathscr{C}$. Hence, we have $\mathscr{C}' \preceq \mathscr{C}.$  

Suppose now that $\mathscr{C}' \prec \mathscr{C}$. Since the set 
    $\{\mathscr{B} \in \text{CAD}^r(\Fr) \; | \; \mathscr{C}' \prec \mathscr{B} \prec \mathscr{C}\}$ is finite, it is sufficient to prove that there exists $\mathscr{D} \in \text{CAD}^r(\Fr)$ such that $\mathscr{C}' \leftarrow \mathscr{D}\preceq \mathscr{C}$. We will build such a CAD $\mathscr{D}$ by adding a suitable section of $\mathscr{C}$ to the sections of $\mathscr{C}'$.
    We will select this section on the $l^{\text{th}}$ level of $\Cr$, where $l \in \{1,\ldots,n\}$ is such that $
        \mathscr{C}'_k = \mathscr{C}_k$ if $k < l$ and $\mathscr{C}_{k}' \prec \mathscr{C}_k$ if $k \geq l$ (see Lemma \ref{lemma:tradOrder}).
        
    By definition, there exists a cell $C'_{J:s} \in \mathscr{C}'_l$ which is a non trivial union of cells of $\mathscr{C}_l$, i.e. we can write $C'_{J:s} = \bigcup_{a \in A} C_{I_a : r_a}$ where $A$ is a finite set with at least two elements. Since $\Cr_{l-1}=\Cr'_{l-1}$, both sides of the equality project onto the same cell $C'_J = C_J \subseteq \mathbb{R}^{l-1}$. Hence, we have $I_a = J$ for all $a \in A$. 
        If $x \in C'_J$, we observe that the set $\{y \in \mathbb{R} \; | \; (x,y) \in C'_{J:s}\}$ is not a singleton. It is then an open interval, which means that $C'_{J:s}$ is a sector. Furthermore, there exists $a\in A$ such that $r_a$ is even and is thus the index of a section of $\mathscr{C}_l$. Thus, we can write $s = 2j+1, r_a = 2m$ for some $j \in \{0,\ldots, u'_J\}, m \in \{1,\ldots,u_J\}$.
        By definition, the sector $C'_{J:2j+1}$ is defined using the functions $\xi'_{J:2j}$ and $\xi'_{J:2(j+1)}$, and $C_{J:2m}$ is the graph of $\xi_{J:2m}$. Since $C_{J:2m} \subseteq C'_{J:2j+1}$, we have $\xi'_{J:2j} < \xi_{J:2m} < \xi'_{J:2(j+1)}$ over $C'_J$.
        Then, we construct the CAD $\mathscr{D} = (\mathscr{D}_1,\ldots,\mathscr{D}_n)$ explicitly by
    \begin{itemize}
        \item setting $\mathscr{D}_k = \mathscr{C}_k$ if $k < l$;
        \item replacing $C'_{J:2j+1}$ in $\mathscr{C}'_l$ by the three cells obtained by slicing it with $\xi_{2m}$ to obtain $\mathscr{D}_l$;
        \item slicing accordingly the cells of $\mathscr{C}'_k$ for $k > l$ above $C'_{J:2j+1}$ to obtain $\mathscr{D}_k$.
    \end{itemize} 
    By construction, there exists a $\text{CAD}^r(\Fr)$ reduction rule $\Phi_{J:2(j+1)}$ from $\mathscr{D}$ to $\mathscr{C}'$. 
    We finally show by induction that $\mathscr{D}_k \preceq \mathscr{C}_k$ for $k \in \{l,\ldots,n\}$.
\end{proof}
\begin{example}\label{ex:hasseDisk}
    We consider the same disk $S$ and CADs $\mathscr{C}$ and $\mathscr{C}'$ as in Example \ref{ex:diskMotiv}. 
    We define the CAD $\mathscr{C}''$ depicted in the top of Figure~\ref{fig:hasse} by slicing the cylinders above $C'_3, C'_4$ and $C'_5$ by means of the function $f$ defined on the interval $(-1,1)$ by $f(x) =  1 - (2(x^2-1))^{-1}$.
    Figure~\ref{fig:hasse} shows the Hasse diagram of elements of $\text{CAD}^r(S)$ smaller than or equal to $\mathscr{C}''$. 
    For instance, $\Phi_{36}$ merges the cells surrounding the third section of the third cylinder.
\end{example}

From Theorem \ref{prop:lien-red-ordre}, we deduce immediately Algorithm \ref{algo:Min} which gives an effective way to build a minimal $\CAD^r$ adapted to $\Fr$ as a post-processing operation once an element of $\text{CAD}^r(\Fr)$ is computed. 

\begin{algorithm}[H]
    \caption{\texttt{Minimal}($\mathscr{C}, \Fr$)}
    \label{algo:Min}
\begin{algorithmic}[1]
    \Require $\mathscr{C} \in \text{CAD}^r(\Fr)$ 
    \Ensure A minimal $\CAD^r$ $\mathcal{M}$ adapted to $\Fr$ such that $\mathcal{M} \preceq \mathscr{C}$ 
    \State Construct the set Red(Tree($\mathscr{C}$)) of reduction rules from Tree($\mathscr{C}$)
    \If{$\exists \Psi \in \text{Red(Tree}(\mathscr{C}))$ s.t. $\Psi$ lifts to a $\CAD^r(\Fr)$ reduction rule $\Phi$ from $\mathscr{C}$ to $\mathscr{C}'$}
    \State \Return \texttt{Minimal}($\mathscr{C}', \Fr$)
    \Else{}
    \State \Return $\mathscr{C}$
    \EndIf
    \end{algorithmic}
\end{algorithm}

We now apply this algorithm to the CADs $\mathscr{C}$ and $\mathscr{C}'$ adapted to the Trousers $\mathbb{T}$ discussed in Theorem \ref{ex:trousers1}. Note that similar considerations hold for the Trousers $\mathbb{T}^\omega$.

\begin{example}\label{ex:trousers2}
    In order to apply  Algorithm \ref{algo:Min} to $(\mathscr{C},\mathbb{T})$ and $(\mathscr{C}',\mathbb{T})$, we compute the sets $\text{Red(Tree}(\mathscr{C}))$ and $\text{Red(Tree}(\mathscr{C}'))$. In this case, we can immediately identify on Figure~\ref{fig:trousers2} the even nodes of these CAD trees satisfying Condition \eqref{eqn:L} and we obtain
    \begin{align*}
        \text{Red(Tree}(\mathscr{C})) = \{\Psi_{12}\} \;\text{ and }\;
        \text{Red(Tree}(\mathscr{C}')) = \{\Psi_{32}\}.
    \end{align*}

    To conclude, it is sufficient to show that neither of these two CAD tree reduction rules lifts to a $\CAD^0(\mathbb{T})$ reduction rule. Suppose that $\Psi_{12}$ lifts to a reduction rule from $\mathscr{C}$ to a CAD $\mathscr{D}$ adapted to $\mathbb{T}$. The section $D_{112}$ of $\mathscr{D}$ is the union of cells of $\mathscr{C}$ whose indices are mapped to $112$ by $\psi_{12}$, which are the sections $C_{112}, C_{122}$ and $C_{132}$. In other words, $D_{112}$ is the graph of a continuous semi-algebraic function defined over the union of $C_{11}, C_{12}$ and $C_{13}$.
This function is obviously not continuous by definition of $\mathbb{T}$. 
Similar considerations show that $\Psi_{32}$ does not lift to a CAD reduction for $\mathscr{C}'$.
 Hence, the CADs returned by \texttt{Minimal}$(\mathscr{C},\mathbb{T})$ and \texttt{Minimal}$(\mathscr{C}',\mathbb{T})$ are respectively $\mathscr{C}$ and $\mathscr{C}'$. This shows that these CADs are minimal elements of $\CAD^0(\mathbb{T})$. 
     \begin{figure}[H]
        \center 
        \includegraphics[scale=0.6]{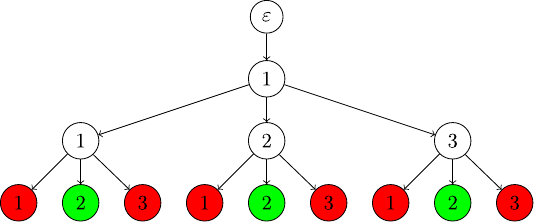} \; \; \; 
        \includegraphics[scale=0.6]{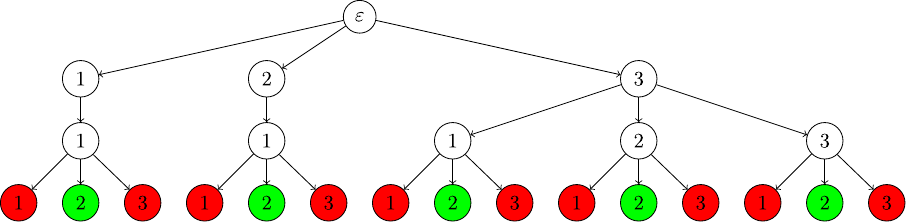}
        \caption{Tree($\mathscr{C}$) and Tree($\mathscr{C}'$)}
        \label{fig:trousers2}
    \end{figure}
\end{example}

\begin{remark}
    In the previous example we just had to show that two particular partitions were not CADs.
    This should be compared with the large number of elements in SSP($\mathscr{C}$) and SSP($\mathscr{C}'$) that would have been processed by direct inspection (see Remark \ref{rem:naive}). 
\end{remark}


\subsection{Representation of the poset \((\mathrm{CAD}^{r}(\Fr), \preceq)\)}
We show in Theorem \ref{prop:TR} that the reduction relation $\leftarrow$ on $\CAD^r(\Fr)$ is the transitive reduction (see \cite{TR1972}) of the infinite poset $(\CAD^r(\Fr), \preceq)$. This phenomenon may already be observed in Example \ref{ex:hasseDisk} where we determined the arrows of a Hasse Diagram corresponding to all $\CAD^0$ adapted to $S$ and smaller than or equal to a given one.

 Recall that relation on $\CAD^r(\Fr)$ is a subset of $\CAD^r(\Fr) \times \CAD^r(\Fr)$ and that the set of relations on $\CAD^r(\Fr)$ is naturally endowed with a partial order given by the inclusion. If $R$ is a relation on $\CAD^r(\Fr)$, we write $\Cr' R \Cr$ when $(\Cr', \Cr) \in R$, and $\Cr' \cancel{R} \Cr$ otherwise. We denote by $R^+$ the transitive closure of $R$, i.e. the smallest transitive relation containing $R$.


\begin{theorem}\label{prop:TR}
    The relation $\leftarrow$ on $\CAD^r(\Fr)$ is the minimum relation whose transitive closure is $\prec$. 
\end{theorem}
\begin{proof}
By Theorem \ref{prop:lien-red-ordre} the transitive closure of $\leftarrow$ is $\prec$. We prove by contradiction that $\leftarrow$ is a minimum in the set of relations on $\CAD^r(\Fr)$ having this property.  Let $R$ be a relation on $\CAD^r(\Fr)$ whose transitive closure $R^+$ is $\prec$ and assume that $\leftarrow$ is not smaller than or equal to $R$. By definition of the order on relations, there exists $\Cr, \Cr' \in \CAD^r(\Fr)$ such that $\Cr' \leftarrow \Cr$ but $\Cr' \cancel{R} \Cr$. By definition of $\leftarrow$, we thus have $\Cr' \prec \Cr$, and therefore $\Cr' R^+ \Cr$, by definition of $R$. Since $\Cr\cancel{R} \Cr'$, there exists $\Dr \in \CAD^r(\Fr)$ such that $\Cr' R^+ \Dr R \Cr$. By assumption on $R$, we know that $\Cr' \prec \Dr \prec \Cr$ and hence $\Cr' \leftarrow^+ \Dr \leftarrow^+ \Cr$. Without loss of generality, we can assume that $\Cr' \leftarrow^+ \Dr \leftarrow \Cr$, or equivalently that $\Cr' \prec \Dr \leftarrow \Cr$. As depicted in the following diagram, we denote by $\Phi_A$ and $\Phi_B$ the reductions defined from $\Cr$ to $\Cr'$ and $\Dr$ respectively. 
\[\begin{tikzcd}
	& \Cr \\
	{\Cr'} & \prec & \Dr
	\arrow["{\Phi_A}"', from=1-2, to=2-1]
	\arrow["{\Phi_B}", from=1-2, to=2-3]
\end{tikzcd}\]
We now show that this situation leads to a contradiction in all the possible configurations for the even indices $A$ and $B$. We first recall that $\Cr' \prec \Dr$ implies $\Cr'_k \preceq \Dr_k$ for all $k \in \{1, \ldots n\}$ (see Lemma \ref{lemma:tradOrder}). Denoting by $|I|$ the length of the tuple $I$, we have the following cases:
\begin{itemize}
    \item If $A=B$, then $\Cr'=\Dr$, which is a contradiction since $\Cr'\prec \Dr$.
    \item If $|A|=|B|$ and $A\neq B$, then the section $C_B$ is a cell of $\Cr'_{|A|}$ since 
    \[\Cr'_{|A|}=\Cr_{|A|}\setminus\{C_{A-e_{|A|}},C_A,C_{A+e_{|A|}}\}\cup\{C_{A-e_{|A|}}\cup C_A\cup C_{A+e_{|A|}}\},\] 
    and $C_{A-e_{|A|}},C_{A+e_{|A|}}$ are sectors. Observe that $\Dr_{|A|}$ is given similarly by 
     \[\Dr_{|A|}=\Cr_{|A|}\setminus\{C_{B-e_{|B|}},C_B,C_{B+e_{|B|}}\}\cup\{C_{B-e_{|B|}}\cup C_B\cup C_{B+e_{|B|}}\},\]
     and that $C_B$ does not meet the cells of $\Cr_{|A|}\setminus\{C_{B-e_{|B|}},C_B,C_{B+e_{|B|}}\}$ and cannot contain $C_{B-e_{|B|}}\cup C_B\cup C_{B+e_{|B|}}$. Hence, $C_B$ is not a union of cells of $\Dr_{|A|}$, in contradiction with $\Cr'_{|A|} \preceq \Dr_{|A|}$. 
     \item If $|A| > |B|$, by definition of the reductions we have $\Cr'_{|B|} = \Cr_{|B|}$. But we also have $\Cr'_{|B|} \preceq \Dr_{|B|} \prec \Cr_{|B|}$, and therefore $\Cr'_{|B|}\prec\Cr_{|B|}$, which is absurd.
     \item If $|A| < |B|$, then the cell $C_B$ of $\Cr_{|B|}$ is a subset of the sector $C_{B-e_{|B|}}\cup C_B\cup C_{B+e_{|B|}}$ of $\Dr_{|B|}$ and it is also, by definition, a subset of the cell $C'_{\psi_A(B)}$ of $\Cr'_{|B|}$. So we have $(C_{B-e_{|B|}}\cup C_B\cup C_{B+e_{|B|}})\cap C'_{\psi_A(B)}\neq \varnothing$ and since $\Cr'\prec\Dr$, $(C_{B-e_{|B|}}\cup C_B\cup C_{B+e_{|B|}})\subseteq  C'_{\psi_A(B)}$. This is impossible since by definition, $C'_{\psi_A(B)}$ is either $C_B$ or $C_{B-e_{|A|}}\cup C_B\cup C_{B+e_{|A|}}$.
\end{itemize}
\end{proof}


\subsection{Minimum CAD and confluence}

\begin{figure}[H]
        \center 
        \includegraphics[scale=0.3]{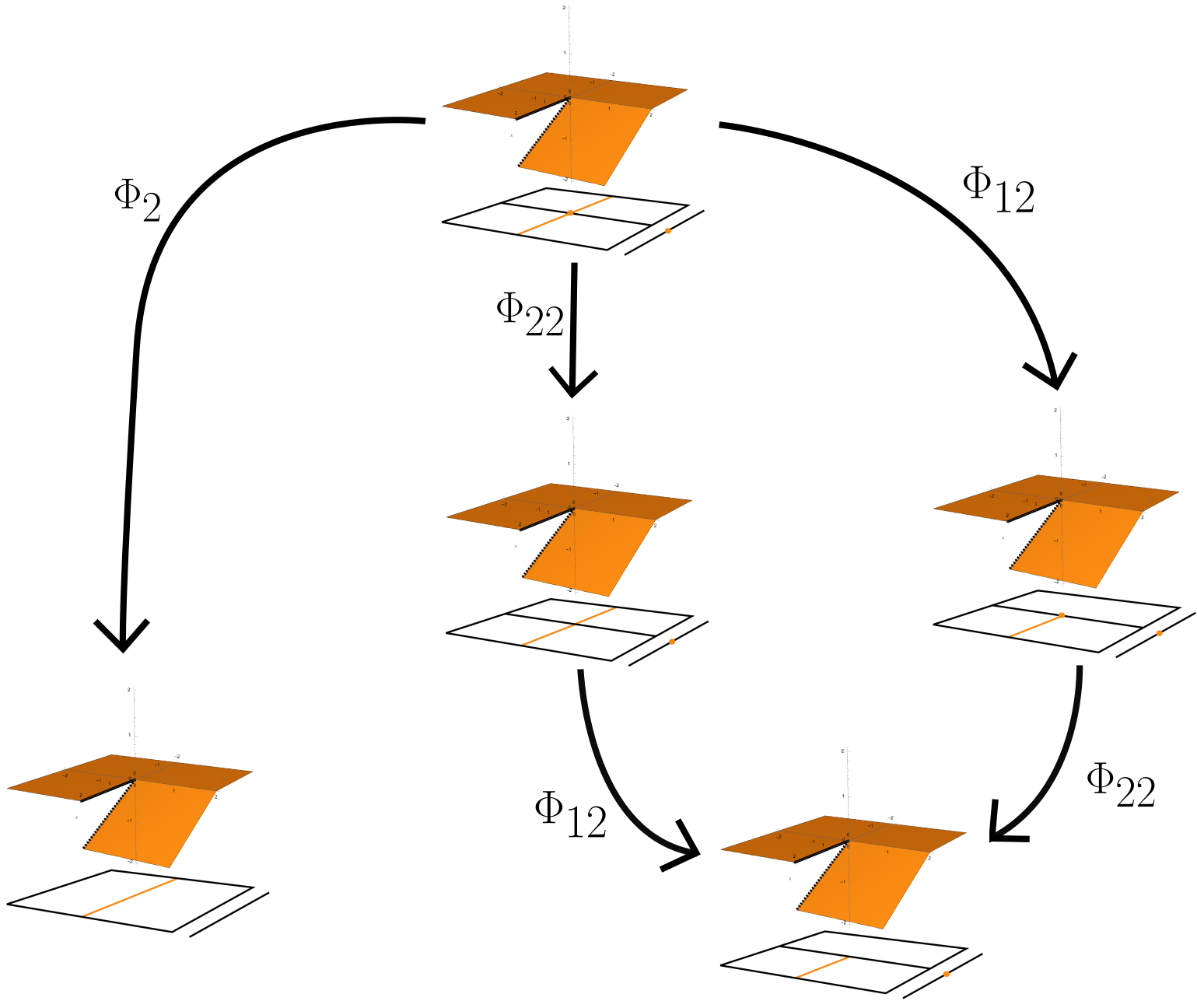}
        \caption{Non-confluence of the Trousers}
        \label{fig:trousers3}
    \end{figure}

In Theorem \ref{ex:trousers1} we proved that $\text{CAD}^0(\mathbb{T})$ has no minimum by considering two distinct minimal adapted CADs $\mathscr{C}$ and $\mathscr{C}'$ of class $C^0$. These CADs can actually be obtained by reductions of a common CAD $\overline{\mathscr{C}} \in \text{CAD}^0(\mathbb{T})$ which is depicted at the top of Figure \ref{fig:trousers3} while $\mathscr{C}$ and $\mathscr{C}'$ are at the bottom.


We thus have $\mathscr{C} \stackrel{*}{\leftarrow} \overline{\mathscr{C}} \stackrel{*}{\to} \mathscr{C}'$ but in view of Theorem~\ref{prop:lien-red-ordre}, $\mathscr{C}$ and $\mathscr{C}'$ cannot be reduced to a common $\CAD^0$. This observation relates the existence of minimum element to the confluence property of $\text{CAD}^0(\mathbb{T})$. We now show that this is a general phenomenon.

We consider $\text{CAD}^r(\Fr)$ as the set of objects of an abstract reduction system (see \cite{TRaAT}) endowed with the reduction relation introduced in Definition \ref{def:CADRed}. Confluence, also known as Church-Rosser or diamond property, appears in various areas of mathematics and computer science. 
We recall its definition in this particular case.

\begin{definition}
    The reduction system $\text{CAD}^r(\Fr)$ is globally confluent if for all $\mathscr{C}, \overline{\mathscr{C}}, \mathscr{C}'  \in \text{CAD}^r(\Fr)$ such that $\mathscr{C} \stackrel{*}{\leftarrow} \overline{\mathscr{C}} \stackrel{*}{\to} \mathscr{C}'$, there exists $\underline{\mathscr{C}} \in \text{CAD}^r(\Fr)$ such that $\mathscr{C} \stackrel{*}{\to} \underline{\mathscr{C}} \stackrel{*}{\leftarrow} \mathscr{C}'$. It is 
        locally confluent if for all $\mathscr{C}, \overline{\mathscr{C}}, \mathscr{C}'  \in \text{CAD}^r(\Fr)$ such that $\mathscr{C} \stackrel{}{\leftarrow} \overline{\mathscr{C}} \stackrel{}{\to} \mathscr{C}'$, there exists $\underline{\mathscr{C}} \in \text{CAD}^r(\Fr) : \mathscr{C} \stackrel{*}{\to} \underline{\mathscr{C}} \stackrel{*}{\leftarrow} \mathscr{C}'$.
    \end{definition}
    
    Observing that there exists no infinite chain of reductions in $\text{CAD}^r(\Fr)$ and applying Newman's Lemma (\cite[Lemma 2.7.2]{TRaAT}) we obtain the following result. 
    \begin{lemma}
        The rewriting system $\text{CAD}^r(\Fr)$ is globally confluent if and only if it is locally confluent.
    \end{lemma}
    In view of this lemma, we say that $\text{CAD}^r(\Fr)$ is confluent if it is globally confluent or locally confluent.
    We can now state and prove the main result of this section which is reminiscent of the characterization of Gröbner bases by means of confluence (see \cite[Chapter~8]{TRaAT}).
    
    \begin{theorem}\label{thrm:confl}
    There exists a minimum in $\text{CAD}^r(\Fr)$ if and only if the reduction system $\text{CAD}^r(\Fr)$ is confluent.
    \end{theorem}
    \begin{proof}         
    Assume that $\text{CAD}^r(\Fr)$ has a minimum $\mathcal{M}$. For every $\mathscr{C},\overline{\mathscr{C}},\mathscr{C}' \in \text{CAD}^r(\Fr)$ such that  $\mathscr{C} \leftarrow \overline{\mathscr{C}} \to \mathscr{C}'$, $\mathcal{M}$ is smaller than or equal to $\mathscr{C}$ and $\mathscr{C}'$. Using Theorem \ref{prop:lien-red-ordre}, this is equivalent to $\mathscr{C} \stackrel{*}{\to} \mathcal{M} \stackrel{*}{\leftarrow} \mathscr{C}'$, so $\text{CAD}^r(\Fr)$ is confluent.
    
    We assume now that $\text{CAD}^r(\Fr)$ is confluent and show that it has a minimum. By Proposition \ref{prop:uniqueMin}, it is sufficient to show that there is at most one minimal element. If $\mathscr{C}$ and $\mathscr{C}'$ are two minimal elements of $\text{CAD}^r(\Fr)$ we obtain a $\CAD^r$ $\overline{\mathscr{C}}$ which is finer than $\mathscr{C}$ and $\mathscr{C}'$ by building a $\CAD^r$ adapted to all the cells of $\mathscr{C}$ and all the cells of $\mathscr{C}'$ (see Remark \ref{rem:Collins}). Hence, we have $\mathscr{C} \stackrel{*}{\leftarrow} \overline{\mathscr{C}} \stackrel{*}{\to} \mathscr{C}.$  By confluence there exists $\underline{\mathscr{C}} \in \text{CAD}^r(\Fr) : \mathscr{C} \stackrel{*}{\to} \underline{\mathscr{C}} \stackrel{*}{\leftarrow} \mathscr{C}'$. By minimality of $\mathscr{C}$ and $\mathscr{C}'$, we have $\mathscr{C} =\underline{\mathscr{C}}= \mathscr{C}'$, as requested.  
    \end{proof}
    


%% file: positive.tex
\section{Existence results for minimum CADs}\label{sec:positive}



We now develop more concrete criteria and provide examples of semi-algebraic sets that do admit a minimum $\CAD^r$.
The main result of this section is a general criterion (Theorem \ref{thrm:criterionP}) that establishes an equivalence between the existence of a minimum in $\CAD^r(\Fr)$ and the existence of a minimum of a poset $\pi_{n-1}(\CAD^r(\Fr))$ (see below) of CADs of $\R^{n-1}$.
However, describing explicitly the set $\pi_{n-1}(\CAD^r(\Fr))$ can be intricate in practice, but as we shall see, it is relevant to seek for general properties that are shared by all its elements, that is to consider overapproximations $\mathcal{O}\supseteq\pi_{n-1}(\CAD^r(\Fr))$ that are easier to handle. 
Next, we determine particular instances of overapproximations and use them to describe explicitly minimum CADs adapted to some semi-algebraic sets. 
These developments could pave the way towards the description of natural classes of (finite families of) semi-algebraic sets of $\R^n$ that do admit a minimum CAD since Theorem \ref{thrm:criterionP} initiates an induction.

\subsection{General criteria}

We now define some posets of $\CAD$s that are naturally associated with $\CAD^r(\Fr)$ which are relevant for the statement of the main criterion.

 \begin{definition}\label{rem:pik}
     For all $k \in \{1, \ldots, n\}$, we denote by $\pi_k : \R^n \to \R^{k}$ the projection defined by $\pi_k(x_1, \ldots, x_n) = (x_1, \ldots, x_k)$ and use the same notation for its natural successive extensions to subsets of $\R^n$, families of subsets of $\R^n$, partitions of $\R^n$ and finally to sets of partitions. 
\end{definition}
In particular, for $\Fr = (S_1, \ldots, S_p)$, we write $\pi_k(\Fr) = (\pi_k(S_1), \ldots, \pi_k(S_p))$ and for every $\CAD$ $\Cr$ of $\R^n$, we have $\pi_k(\Cr) = \Cr_k$ and therefore
    \[\pi_k(\CAD^r(\Fr)) = \{\Cr_{k} \; | \; \Cr \in \CAD^r(\Fr)\}.\]
However, $\pi_k(\CAD^r(\Fr))$ is in general not equal to $\CAD^r(\pi_k(\Fr))$ (see Proposition \ref{lemma:projk}), and it might not be equal to a poset of the form $\CAD^r(\Fr_k)$, with $\Fr_k$ a finite family of semi-algebraic sets of $\R^{k}$ (see Proposition \ref{cor:caractFk}). It is nevertheless endowed with the partial order $\preceq$ given by refinement (see Definition \ref{def:order}), since it is a subset of the set of partitions of $\R^{k}$.

\begin{lemma}\label{lemma:criterionPk}
    If the poset $(\CAD^r(\Fr), \preceq)$ has a minimum element $\Mr$, then $\Mr_k$ is a minimum element of the poset $(\pi_k(\CAD^r(\Fr)), \preceq)$ for every $k \in \{1, \ldots, n-1\}$.
\end{lemma}
\begin{proof}
    By definition, if $\Dr \in \pi_k(\CAD^r(\Fr))$, then there exists $\Cr \in \CAD^r(\Fr)$ such that $\Cr_{k} = \Dr$. In particular, $\Mr \preceq \Cr$ and thus $\Mr_{k} \preceq \Cr_{k}=\Dr$ by Lemma \ref{lemma:tradOrder}. 
\end{proof}
The converse of this lemma would allow to infer the existence of a minimum element in $(\CAD^r(\Fr), \preceq)$ from that of minimum elements in the posets $(\pi_k(\CAD^r(\Fr)), \preceq)$. The following theorem provides a stronger statement using only $(\pi_{n-1}(\CAD^r(\Fr)), \preceq)$.



\begin{theorem}\label{thrm:criterionP}
    Let $\mathcal{O}$ be a set of partitions of $\R^{n-1}$ such that $\pi_{n-1}(\CAD^r(\Fr))\subseteq \mathcal{O}$. Assume that $\Mr$ a minimal $\CAD^r$ adapted to $\Fr$ such that $\Mr_{n-1}$ is a minimum of $\mathcal{O}$. Then $\Mr$ is a minimum $\CAD^r$ adapted to $\Fr$. 
    In particular, the poset $(\CAD^r(\Fr), \preceq)$ has a minimum element if and only if the poset $(\pi_{n-1}(\CAD^r(\Fr)), \preceq)$ has a minimum element.
\end{theorem}
\begin{proof}
   By Proposition \ref{prop:uniqueMin}, it is sufficient to prove that $\Mr$ is the unique minimal element in $\CAD^r(\Fr)$, i.e. that if $\Cr$ is a minimal element in $\CAD^r(\Fr)$, then $\Cr=\Mr$.   We show that $\Mr \preceq \Cr$ and we conclude by minimality of $\Cr$.
    

    Consider a cell $M_{I:a}$ of $\Mr$ and observe that $M_I \in \Mr_{n-1}$. By assumption, $\Cr_{n-1} \in \mathcal{O}$ and $\Mr_{n-1}$ is a minimum of $\mathcal{O}$. This implies that $\Mr_{n-1} \preceq \Cr_{n-1}$, and thus the existence of $J_1, \ldots, J_K\in (\N^*)^{n-1}$ such that $M_I = \cup_{k=1}^K C_{J_k}$, where the $C_{J_k}$'s are cells of $\Cr_{n-1}$. Finally, by Corollary \ref{prop:outil}, for all $k \in \{1,\ldots, K\}$, the odd number of cells of $\Mr$ in the cylinder above $M_{I}$ is the same as the number of cells of $\Cr$ in the cylinder above $C_{J_k}$, but also that $M_{I:a} = \cup_{k=1}^K C_{J_k : a}$, hence $\Mr \preceq \Cr$.   

    For the particular case, we denote by $\mathcal{N}$ the minimum of the poset $(\pi_{n-1}(\CAD^r(\Fr)), \preceq)$. By definition, there exists $\Cr \in \CAD^r(\Fr)$ such that $\Cr_{n-1} = \mathcal{N}.$ Consider $\Mr$ a minimal $\CAD^r$ of $\Fr$ such that $\Mr \preceq \Cr$. By Lemma \ref{lemma:tradOrder}, we have $\Mr_{n-1} \preceq \Cr_{n-1} = \mathcal{N}$. Since $\mathcal{N}$ is a minimum of $(\pi_{n-1}(\CAD^r(\Fr)), \preceq)$, we obtain that $\Mr_{n-1} = \mathcal{N}$. We conclude by considering $\mathcal{O} = \pi_{n-1}(\CAD^r(\Fr))$.
\end{proof}
We now deduce a new theoretical characterisation of the fact that the poset $\CAD^r(\Fr)$ admits a minimum. 
A subset $P$ of $\CAD^r(\R^n)$ is upward closed (with respect to $\preceq$) if for all $\Cr \in P$ and all $\Dr \in \CAD^r(\R^n)$ such that $\Cr \preceq \Dr$, we have $\Dr \in P$.  
\begin{lemma}\label{lemma:upwardclosed}
For $k\in \{1, \ldots, n\}$, the subset $\pi_k(\CAD^r(\Fr))$ of $\CAD^r(\R^k)$ is upward closed.
\end{lemma}
\begin{proof}
    Let $\widetilde{\Cr} \in \pi_k(\CAD^r(\Fr))$ and $\Dr \in \CAD^r(\R^k)$ such that $\widetilde{\Cr} \preceq \Dr$. By definition, there exists $\Cr \in \CAD^r(\Fr)$ such that $\Cr_k = \widetilde{\Cr}$. To show that $\Dr \in \pi_k(\CAD^r(\Fr))$, we build level by level an element $\Er \in \CAD^r(\Fr)$ such that $\Er_k = \Dr$ as follows. First, we set $\Er_k = \Dr$. Then, for all $l \in \{k+1, \ldots, n\}$, if the CAD $\Er_{l-1}$ is already built, then the cells of $\Er_l$ are precisely the intersections of the cells of $\Cr_l$ with the cylinders $E \times \R$ for $E \in \Er_{l-1}$. It is clear that $\Er$ is adapted to $\Fr$ since $\Er$ is greater than or equal to $\Cr$, which is adapted to $\Fr$.
\end{proof}

\begin{lemma}\label{lemma:subposetP}
Let $P$ be a subset of $\CAD^r(\R^n)$. If $P$ is upward closed and admits a minimum $\Mr$, then $P = \CAD^r(\Mr)$. If there exists $\Mr \in \CAD^r(\R^n)$ such that $P = \CAD^r(\Mr)$, then $\Mr$ is the minimum of $P$ and $P$ is upward closed.
\end{lemma}
\begin{proof}
    This is a direct consequence of the fact that for all CADs $\Cr$ and $\Dr$ of $\R^n$ of class $C^r$, we have $\Cr \preceq \Dr$ if and only if $\Dr \in \CAD^r(\Cr)$.
\end{proof}

\begin{proposition}\label{cor:caractFk}
    The poset $\CAD^r(\Fr)$ has a minimum if and only if for all $k \in \{1, \ldots, n\}$, there exists a finite family $\Fr_k$ of semi-algebraic sets of $\R^k$ such that \[\pi_k(\CAD^r(\Fr)) = \CAD^r(\Fr_k).\]
\end{proposition}
\begin{proof}
    If $\CAD^r(\Fr)$ has a minimum $\Mr$, then $\Mr_k$ is a minimum of $\pi_k(\CAD^r(\Fr))$ for every $k\in \{1, \ldots, n\}$ (see Lemma \ref{lemma:criterionPk}). Since these posets are all upward closed by Lemma \ref{lemma:upwardclosed}, we obtain that $\pi_k(\CAD^r(\Fr)) = \CAD^r(\Mr_k)$ via Lemma \ref{lemma:subposetP}. 


    For the converse, we proceed by induction on $k \in \{1, \ldots, n\}$ to show that the poset $\pi_{k}(\CAD^r(\Fr))$ admits a minimum.  
    The poset $\pi_1(\CAD^r(\Fr))$ coincides with $\CAD^r(\Fr_1)$ by assumption and it therefore admits a minimum by Theorem \ref{thrm:existenceMinimum}, as the ambient space is $\R$. 
    Let us assume that $\pi_{k-1}(\CAD^r(\Fr))$ admits a minimum, for a given $k \in \{2, \ldots, n\}$ and show that $\pi_{k}(\CAD^r(\Fr))$ admits a minimum, by applying Theorem \ref{thrm:criterionP}.
    By assumption, we can write $\pi_k(\CAD^r(\Fr))=\CAD^r(\Fr_k)$ and moreover we have
    \[\pi_{k-1}(\CAD^r(\Fr_k))= \pi_{k-1}(\pi_k(\CAD^r(\Fr)))=\pi_{k-1}(\CAD^r(\Fr)).\] 
    The latter admits a minimum, hence Theorem \ref{thrm:criterionP} ensures that $\CAD^r(\Fr_k)=\pi_k(\CAD^r(\Fr))$ has a minimum.
\end{proof}

To illustrate the previous result, consider the Trousers $\mathbb{T}$ from Theorem \ref{ex:trousers1} and the CAD $\Cr$ from the corresponding proof. We know that $\pi_1(\CAD^0(\mathbb{T}))$ is upward closed and contains the trivial CAD  $\Cr_1 = \{\R\}$ of $\R$, hence we have $\pi_1(\CAD^0(\mathbb{T})) = \CAD^0(\R)$.  Moreover, since the poset $\pi_3(\CAD^0(\mathbb{T})) = \CAD^0(\mathbb{T})$ does not admit a minimum element, Proposition \ref{cor:caractFk} implies that there does not exist a finite family $\Fr_2$ of semi-algebraic sets of $\R^2$ such that $\pi_2(\CAD^0(\mathbb{T})) = \CAD^0(\Fr_2).$

\subsection{Applications} 

In this subsection we define explicitly suitable overapproximations $\mathcal{O}$ of $\pi_{n-1}(\CAD^r(\Fr))$ and apply Theorem \ref{thrm:criterionP} to present various explicit examples of (finite families of) semi-algebraic sets of $\R^3$ which do admit a minimum CAD. Similar constructions extend to $\R^n$.

\begin{proposition}\label{lemma:projk} 
    For every $k \in \{1, \ldots, n\}$, we have $\pi_k(\CAD^r(\Fr)) \subseteq \CAD^r(\pi_k(\Fr))$. 
\end{proposition}
The other inclusion is not true in general (see for instance Example \ref{ex:doubleparabolas}).
\begin{proof}
    We show that if $\Cr \in \CAD^r(\Fr)$, then we have $\Cr_{k} \in \CAD^r(\pi_k(\Fr))$. By assumption, $S_j = \bigcup \{ C \; | \; C \in \Cr, C \subseteq S_j \}$ for all $S_j \in \Fr$ and hence
    $\pi_k(S_j) = \bigcup \{ \pi_k(C) \; | \;C \in \Cr, C \subseteq S_j \}.$
    The result follows since the cells of $\Cr_k$ are precisely the sets $\pi_k(C)$ with $C \in \Cr$.
\end{proof}

\begin{example}\label{ex:closedball} 
    The closed ball $\mathbb{B}$ of $\R^3$ centred at the origin and of radius one admits a minimum $\CAD^r$. We show this using the criterion (Theorem \ref{thrm:criterionP}) with the overapproximation $\mathcal{O} = \CAD^r(\pi_2(\mathbb{B}))$. Observe that $\pi_2(\mathbb{B})$ is the closed disk of the plane centred at the origin and of radius one.  
    We consider $\Mr \in \CAD^r(\mathbb{B})$ whose $\Mr_1$ and $\Mr_2$ are exactly the CADs $\Cr_1$ and $\Cr_2$ described in Example \ref{ex:diskMotiv}. The cells in $\R^3$ of $\Mr$ are obtained by splitting exactly the cylinders built above the cells of $\Mr_2$ by the the roots of the polynomial $x_1^2 + x_2^2 + x_3^2 -1$ with respect to $x_3$.
    The CADs $\Mr$ and $\Mr_2$ are respectively minimal elements of $\CAD^r(\mathbb{B})$ and of $\mathcal{O}$ since we can show that there exists no reduction rule defined from their respective CAD tree. Since $\pi_2(\mathbb{B})$ is a semi-algebraic set of $\R^2$, Theorem \ref{thrm:existenceMinimum} implies that $\Mr_2$ is actually a minimum of $\mathcal{O}$. We can thus apply the criterion, obtaining that $\Mr$ is the minimum $\CAD^r$ of $\mathbb{B}$. 
\end{example}



    From Lemma \ref{lemma:compl-proj}, we already know that the semi-algebraic set $S \equiv x_1^2 + x_2^2 + x_3^2 > 1$ of $\R^3$, which is the complement of $\mathbb{B}$, admits a minimum $\CAD^r$.
    However, notice that since $\pi_2(S) = \R^2$, we cannot apply Theorem \ref{thrm:criterionP} with the the overapproximation $\CAD^r(\pi_2(S))$ of $\pi_2(\CAD^r(S))$ as  in Example \ref{ex:closedball}. Indeed, there obviously does not exist any $\Mr \in \CAD^r(S)$ whose  projection $\Mr_2$ is the minimum of $\CAD^r(\pi_2(S))$, which is $\{\R^2\}$. 
    This suggests to search for other overapproximations in order to apply Theorem \ref{thrm:criterionP} to other classes of examples.

For this purpose, we consider the collection $\text{CC}(\Fr)$ of connected components of the semi-algebraic sets $S_1, \ldots, S_p$. 
It is known that 
 $\text{CC}(\Fr)$ is a finite family of semi-algebraic sets of $\R^n$ (see for instance Theorem 5.21 of \cite{basu2007}), and which we may suppose to be numbered once and for all. 

\begin{proposition}\label{prop:CC}
    We have $\CAD^r(\Fr) = \CAD^r(\text{CC}(\Fr)).$
\end{proposition}
\begin{proof}
    Consider $\Cr \in \CAD^r(\Fr)$, $C \in \Cr$ and $S_i'$ a connected component of some $S_i \in \Fr$ and assume that $C\cap S_i'\neq\varnothing$. On the one hand,  $C \subseteq S_i$ by assumption on $\Cr$. On the other hand, since all CAD cells are connected, we obtain that $S_i'\cup C$ is a connected subset of $S_i$. 
    By definition of $S_i'$, we must have $C\subseteq S_i'$. Therefore $\Cr$ is adapted to any of the connected components of $S_i$, again by Lemma \ref{lemma:compl-proj}. 
    The other inclusion is direct, since any semi-algebraic set is the union of its connected components.
%
\end{proof}


    

\begin{example}\label{ex:doubleparabolas}
    We show that the semi-algebraic set $\mathbb{P}$ defined by
    \[\mathbb{P} = \left\{(x,y,z) \in \R^3 \; | \; \left(x =- 2 (z - \frac{1}{2})^2 \right) \lor \left(x = 2 (z + \frac{1}{2})^2\right) \right\} \]
    and depicted in Figure \ref{fig:2parabolas3D} admits a minimum $\CAD^r$. 
    The set $\mathbb{P}$ admits two connected components, which are the sets $\mathbb{P}_1 \equiv x =- 2 (z - \frac{1}{2})^2$ and $\mathbb{P}_2 \equiv x = 2 (z + \frac{1}{2})^2$. 
    Applying Proposition \ref{prop:CC}, we know that $\CAD^r(\mathbb{P}) \subseteq \CAD^r(\mathbb{P}_1)$. Combining this with Lemma \ref{lemma:projk} and with the fact that $\pi_2(\mathbb{P}_1) = (-\infty;0] \times \R$, we obtain 
    \[
    \pi_2(\CAD^r(\mathbb{P})) \subseteq \pi_2(\CAD^r(\mathbb{P}_1)) \subseteq  \CAD^r(\pi_2(\mathbb{P}_1)) =  \CAD^r((-\infty;0] \times \R)
    \]
   and the set  $\mathcal{O} = \CAD^r((-\infty;0] \times \R) $ is an overapproximation of $\pi_2(\CAD^r(\mathbb{P}))$. A minimum of $\mathcal{O}$ is given by the CAD $\Mr_2$ defined by the three cells $M_{11} = (-\infty; 0) \times \R, M_{12} = \{0\}  \times \R$ and $M_{13} = (0;+\infty)  \times \R$. A minimal element $\Mr$ of $\CAD^r(\mathbb{P})$ is defined above $\Mr_2$ exactly by the three pairs of functions given by $\pm (\frac{-x}{2})^{\frac{1}{2}} + \frac{1}{2}$ on $M_{11}$, by $\pm \frac{1}{2}$ on $M_{12}$ and by $\pm (\frac{x}{2})^{\frac{1}{2}} - \frac{1}{2}$ on $M_{13}$. By Theorem \ref{thrm:criterionP}, $\Mr$ is a minimum of $\CAD^r(\mathbb{P}).$

    \begin{figure}[H]
    \center
    \includegraphics[scale=0.35]{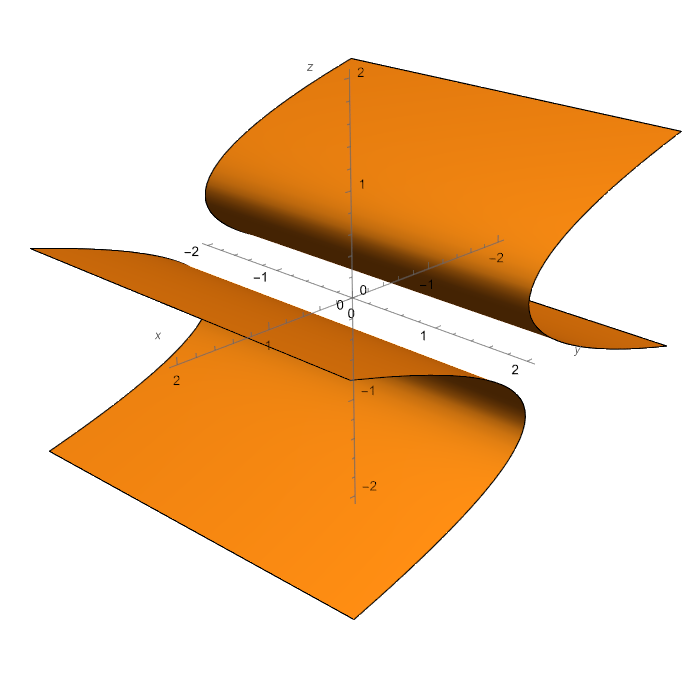}
    \caption{The semi-algebraic set $\mathbb{P}$ with two connected components}
    \label{fig:2parabolas3D}
\end{figure}
\end{example}

%% file: behaviour.tex
In order to introduce the last overapproximation we consider, we start with the following simple example.
\begin{example}\label{ex:halfspace0} 
 Let $S\subseteq \R^2$ be the closed half-plane of equation $y \geq 0$ with the origin $(0,0)$ removed. We first show that if $\Cr$ is a minimal $\CAD^r$ adapted to $S$, then $\Cr_1$ must contain the section $\{0\}$. Suppose the contrary, then the cell $C_i$ of $\Cr_1$ containing $0$ is a sector. By Proposition \ref{rem:top-caract}, there exists a unique section of $\Cr$ above $C_i$, which is exactly $C_i \times \{0\}$. Indeed if $\xi_{i:2},\ldots,\xi_{i:2u_i}$ define the sections over $C_i$, we must have $\{\xi_{i:2j}(x)  \; | \; j \in \{1,\ldots,u_i\}\}=\partial S_x=\{0\}$ for every $x \in C_i$. This is a contradiction since the section $C_i\times\{0\}$ intersects $S$ and $S^c$. Therefore $\Cr_1$ must be adapted to $\{0\}$. 
\end{example}
We observe that in the previous example we have $\partial S_x=\{0\}$ for every $x\in\R$. This does not single out the particular point $0 \in \R$. But we have $S_{x} = [0;+\infty)$ for all $x \neq 0$, whereas $S_{0} = (0; +\infty)$. The former contains its boundary, while the latter does not. This piece of topological information is precisely encoded in the tree associated with the minimum CAD adapted to $S_x$. 
In view of the previous example and Theorem \ref{thrm:existenceMinimum}, we introduce the following definition. 
\begin{definition}\label{def:behaviour1n}
     For every $x \in \R^{n-1}$, the behaviour $[\Fr_x]$ of $\Fr$ above $x$ is the CAD tree associated with the minimum CAD adapted to $\Fr_x = \left((S_1)_x, \ldots, (S_p)_x\right)$.
    The set of all behaviours is denoted by $\mathfrak{B}$.  Finally, for all $\mathfrak{b} \in \mathfrak{B}$, we set  $B(\Fr, \mathfrak{b})=\{x \in \R^{n-1} \; | \; [\Fr_x] = \mathfrak{b}\}.$ 
    \end{definition}
Note that the regularity of the minimum CAD is omitted since the poset $\CAD^r(\Fr_x)$ is the same for all $r \in \N \cup \{\infty, \omega\}$.
  Moreover, the behaviour $[\Fr_x]$, as defined, is a CAD tree $(T^x,L^x)$ of depth $1$ and has exactly $2u_\varepsilon^x + 1$ leaves (see Definitions \ref{def:cad}  and \ref{def:CADTREE}). It is thus completely determined by the list $\left(L^x(1), L^x(2), \ldots, L^x(2u_\varepsilon^x + 1)\right)$ of $p$-tuples in $\{0,1\}^p$. 
\begin{example}\label{ex:behaviour1}
    Let $\Fr = \left(S_1, S_2\right)$ where $S_1 = [-2,0) \cup \{1\}, S_2 = [0,+\infty)$. By Remark \ref{rem:top-caract-bis}, the set of sections of the minimum CAD $\Cr$ adapted to $\Fr$ is exactly given by $\{-2,0,1\}$, which contains $u_\varepsilon = 3$ points, as depicted in Figure \ref{fig:behaviour1D}. Hence, the behaviour of $\Fr$ (above the unique point $0$ of $\R^0$) is the $7$-tuple given by
    \[\left(
    (0, 0),
    (1, 0),
    (1, 0),
    (0, 1),
    (0, 1),
    (1, 1),
    (0, 1)\right).\]
    For instance, the third term is $(1, 0)$ since $C_3 \subseteq S_1$ and $C_3 \subseteq S_2^c$.

        \begin{figure}[H]
    \center

                \begin{tikzpicture}[scale=1,line cap=round,line join=round,x=1.0cm,y=1.5cm,>=latex]
                    \clip (-4,-4.5) rectangle (3.5,-1.3);

                    \filldraw[black] (-1,-4) circle (0pt) node[below]{\small $C_3$};
                    
                    \draw[->] (-3,-4) node[left]{$\Cr$} --(3,-4) node[below]{$\R$};
                    \draw[->] (-3,-3) node[left,blue]{$S_2$}--(3,-3) node[below]{$\R$};
                    \draw[->] (-3,-2) node[left,red]{$S_1$} --(3,-2) node[below]{$\R$};

                    \draw[ dashed] (-2,-2.1)  --(-2,-3.6) ;
                    \draw[ dashed] (0,-2.1)  --(0,-3.6) ;
                    \draw[ dashed] (1,-2.1)  --(1,-3.6) ;

                    \draw[red,very thick] (-2,-2)--(-0.09,-2);
                    \filldraw[red] (-2,-2) circle (2.6pt);
                    \draw[red, very thick](0,-2) circle (2.6pt);
                    \filldraw[red] (1,-2) circle (2.6pt) ;

                    \draw[blue,very thick] (0,-3)--(2.85,-3);
                    \filldraw[blue] (0,-3) circle (2.6pt) ;
                    

                    \filldraw[black] (-2.7,-4) circle (0pt) node[below]{\small $C_1$};
                    \filldraw[black] (-2,-4) circle (2pt) node[below]{\small $C_2$} node[above]{\small $-2$};;
                    \filldraw[black] (0,-4) circle (2pt) node[below]{\small $C_4$} node[above]{\small $0$};
                    \filldraw[black] (0.5,-4) circle (0pt) node[below]{\small $C_5$};
                    \filldraw[black] (1,-4) circle (2pt) node[below]{\small $C_6$} node[above]{\small $1$};
                    \filldraw[black] (2,-4) circle (0pt) node[below]{\small $C_7$};
                \end{tikzpicture}
    \caption{The family $\Fr = (S_1,S_2)$ 
    and its minimum CAD $\Cr$}
    \label{fig:behaviour1D}
\end{figure}
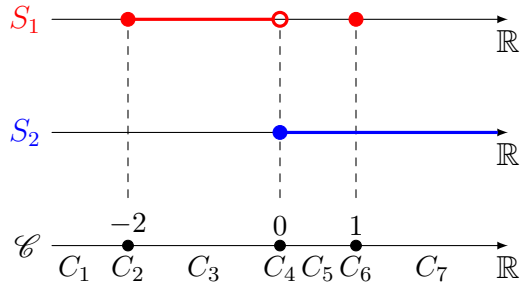
\end{example}


In Example \ref{ex:halfspace0}, we actually showed that $B(S,(0,1,1)) = \R \setminus \{0\}$ and $B(S,(0,0,1)) = \{0\}$.      We also showed that if $\Cr$ is a minimal $\CAD^r$ adapted to $S$, then $\Cr_1$ must be adapted to $\{B(S,(0,1,1)),B(S,(0,0,1))\}$. More generally, we have the following results which enable us to present the desired overapproximation.

\begin{lemma}\label{lemma:f_C}
    Let $\Dr \in \pi_{n-1}(\CAD^0(\Fr))$ and $D$ a cell of $\Dr$. The behaviour $[\Fr_x]$ of $\Fr$ above $x$ is the same for all $x\in D$.
\end{lemma}
\begin{proof}
For every $x\in D$, let $\Mr^{x}$ be the minimum $\CAD$ adapted to $\Fr_x$. The behaviour of $\Fr$ above $x$ is thus $\Tree(\Mr^x,\Fr_x)$. It is characterized by the number $2u_\varepsilon^x + 1$ of leaves and the corresponding list $\left(L^x(1), L^x(2), \ldots, L^x(2u_\varepsilon^x + 1)\right)$. We show that these are independent of $x\in D$. By Proposition \ref{rem:top-caract} and Remark \ref{rem:top-caract-bis}, the sections of $\Mr^x$ are the elements of $\cup_{i=1}^p \partial (S_i)_x$, which are in turn characterized by the sections of any minimal CAD adapted to $\Fr$. 
But by definition there exists $\Er \in \CAD^0(\Fr)$ such that $\Er_{n-1} = \Dr$. By Proposition \ref{prop:existMin}, there exists a minimal element $\Cr$ of $\CAD^0(\Fr)$ such that $\Cr \preceq \Er$ and Lemma \ref{lemma:tradOrder} then asserts that $\Cr_{n-1} \preceq \Er_{n-1} = \Dr$.
By definition, there exists a cell $C_I\in \Cr_{n-1}$ such that $D \subseteq C_I$. Then Proposition \ref{rem:top-caract} (see also Remark \ref{rem:top-caract-bis}) guarantees that $\Mr^{x}$ contains exactly $2u_I+1$ cells. In particular, $u_\varepsilon^x = u_I$ for every $x \in D$. 
    Finally, we show that $L^x(k)$ is independent of $x$ for every $k\in\{1,\ldots,2u_I+1\}$. If we denote by $M^{x}_k$ the $k^{th}$ cell of $\Mr^{x}$, then we have $\{x\}\times M^{x}_k\subseteq C_{I:k}$. The $i^\text{th}$ component of $L^{x}(k)$ is equal to $1$ if and only if $M^{x}_k \subseteq (S_i)_x$ (see Definition \ref{def:CADTREE}), which happens if and only if $C_{I:k} \subset S_i$. The conclusion follows since the latter condition is independent of the choice of $x$ in $D$.
\end{proof}

\begin{proposition}\label{prop:overBehaviour}
    Suppose that $n \geq 2$. 
    The set $\{B(\Fr, \mathfrak{b}) \; | \; \mathfrak{b} \in \mathfrak{B}\}$ is a finite semi-algebraic partition of $\R^{n-1}$, and
    $\pi_{n-1}(\CAD^r(\Fr))\subseteq \CAD^r(B(\Fr, \mathfrak{b}) \; | \; \mathfrak{b} \in \mathfrak{B})$.
\end{proposition}
\begin{proof}
    Let $\Dr \in \pi_{n-1}(\CAD^r(\Fr))$ and $\mathfrak{b}\in\mathfrak{B}$. Since $\Dr$ is a partition of $\R^{n-1}$, for every $x\in B(\Fr,\mathfrak{b})$, there exists $D\in \Dr$, such that $x\in D$. By Lemma \ref{lemma:f_C}, this implies $D \subseteq B(\Fr,\mathfrak{b})$, and therefore $B(\Fr,\mathfrak{b})$ is a (finite) union of cells of $\Dr$. It is therefore semi-algebraic. Moreover, by definition we have $B(\Fr,\mathfrak{b})\cap B(\Fr,\mathfrak{b}')=\varnothing$ when $\mathfrak{b}\neq\mathfrak{b}'$ and $x\in B(\Fr,[\Fr_x])$ for every $x\in\R^{n-1}$. This shows that $\{B(\Fr, \mathfrak{b}) \; | \; \mathfrak{b} \in \mathfrak{B}\}$ is a partition of $\R^{n-1}$, of which $\Dr$ is a refinement. It is finite since $\Dr$ is, and $\Dr$ is adapted to $B(\Fr, \mathfrak{b})$ for every $\mathfrak{b}\in \mathfrak{B}$.
\end{proof}

\begin{example}
Consider the Pointless Ball $\text{PB}$ defined by
\[\text{PB} = \left\{(x,y,z) \in \R^3 \; | \; x^2 + y^2 + z^2 \leq 1 \land z < 1\right\}.\]
The partition described in Proposition \ref{prop:overBehaviour} is given by
\begin{align*}
    B(\text{PB},\mathfrak{b}) \equiv \begin{cases}
        x^2 + y^2 > 1 &\text{ if } \mathfrak{b} = (0),\\
        x^2 + y^2 =1 &\text{ if } \mathfrak{b}  = (0,1,0),\\  
   0<x^2 + y^2 <1 &\text{ if } \mathfrak{b}  = (0,1,1,1,0),\\
   x^2+y^2=0 &\text{ if } \mathfrak{b}  = (0,1,1,0,0),
    \end{cases}
\end{align*}
and provides explicitly the overapproximation $\CAD^r(B(\text{PB}, \mathfrak{b}) \; | \; \mathfrak{b} \in \mathfrak{B})$ of $\pi_2(\CAD^r(\text{PB}))$. The minimum element $\Mr_2$ of this overapproximation is the refinement of the CAD $\Cr'$ of Example \ref{ex:diskMotiv} with the unique additional section given by $\{(0,0)\}$. We construct a CAD $\Mr \in \CAD^r(\text{PB})$ by slicing the cylinders above the cells of $\Mr_2$ with the roots of $x^2 + y^2 + z^2 - 1$ with respect to $z$. Then, we observe that $\Mr$ is minimal since there does not exist any tree reduction defined on its associated tree. It is thus a minimum by Theorem \ref{thrm:criterionP}.


\end{example}

%% file: cas-refs.bib
@ARTICLE{Fortunato2010,
  author  = {Fortunato, S.},
  title   = {Community detection in graphs},
  journal = {Phys. Rep.-Rev. Sec. Phys. Lett.}, 
  volume  = {486},
  year    = {2010},
  pages   = {75-174}
}

@ARTICLE{NewmanGirvan2004,
  author  = {Newman, M. E. J. and Girvan, M.},
  title   = {Finding and evaluating community structure in networks},
  journal = {Phys. Rev. E.}, 
  volume  = {69},
  year    = {2004},
  pages   = {026113}
}

@ARTICLE{Vehlowetal2013,
  author  = {Vehlow, C. and Reinhardt, T. and Weiskopf, D.},
  title   = {Visualizing Fuzzy Overlapping Communities in Networks},
  journal = {IEEE Trans. Vis. Comput. Graph.}, 
  volume  = {19},
  year    = {2013},
  pages   = {2486-2495}
}

@ARTICLE{Raghavanetal2007,
  author  = {Raghavan, U. and Albert, R. and Kumara, S.},
  title   = {Near linear time algorithm to detect community structures in large-scale networks},
  journal = {Phys. Rev E.}, 
  volume  = {76},
  year    = {2007},
  pages   = {036106}
}

@ARTICLE{SubeljBajec2011a,
  author  = {\v{S}ubelj, L. and Bajec, M.},
  title   = {Robust network community detection using balanced propagation},
  journal = {Eur. Phys. J. B.}, 
  volume  = {81},
  year    = {2011},
  pages   = {353-362}
}

@ARTICLE{Louetal2013,
  author  = {Lou, H. and Li, S. and Zhao, Y.},
  title   = {Detecting community structure using label propagation with weighted coherent neighborhood propinquity},
  journal = {Physica A.}, 
  volume  = {392},
  year    = {2013},
  pages   = {3095-3105}
}

@ARTICLE{Clausetetal2004,
  author  = {Clauset, A. and Newman, M. E. J. and Moore, C.},
  title   = {Finding community structure in very large networks},
  journal = {Phys. Rev. E.}, 
  volume  = {70},
  year    = {2004},
  pages   = {066111}
}

@ARTICLE{Blondeletal2008,
  author  = {Blondel, V. D. and Guillaume, J. L. and Lambiotte, R. and Lefebvre, E.},
  title   = {Fast unfolding of communities in large networks},
  journal = {J. Stat. Mech.-Theory Exp.}, 
  volume  = {2008},
  year    = {2008},
  pages   = {P10008}
}

@ARTICLE{SobolevskyCampari2014,
  author  = {Sobolevsky, S. and Campari, R.},
  title   = {General optimization technique for high-quality community detection in complex networks},
  journal = {Phys. Rev. E.}, 
  volume  = {90},
  year    = {2014},
  pages   = {012811}
}

@ARTICLE{FortunatoBarthelemy2007,
  author  = {Fortunato, S. and Barthelemy, M.},
  title   = {Resolution limit in community detection},
  journal = {Proc. Natl. Acad. Sci. U. S. A.}, 
  volume  = {104},
  year    = {2007},
  pages   = {36-41}
}

@ARTICLE{SubeljBajec2011b,
  author  = {\v{S}ubelj, L. and Bajec, M.},
  title   = {Unfolding communities in large complex networks: Combining defensive and offensive label propagation for core extraction},
  journal = {Phys. Rev. E.}, 
  volume  = {83},
  year    = {2011},
  pages   = {036103}
}

@ARTICLE{WangLi2013,
  author  = {Wang, X. and Li, J.},
  title   = {Detecting communities by the core-vertex and intimate degree in complex networks},
  journal = {Physica A.}, 
  volume  = {392},
  year    = {2013},
  pages   = {2555-2563}
}

@ARTICLE{Lietal2013,
  author  = {Li, J. and Wang, X. and Eustace, J.},
  title   = {Detecting overlapping communities by seed community in weighted complex networks},
  journal = {Physica A.}, 
  volume  = {392},
  year    = {2013},
  pages   = {6125-6134}
}

@ARTICLE{Fabioetal2013,
  author  = {Fabio, D. R. and Fabio, D. and Carlo, P.},
  title   = {Profiling core-periphery network structure by random walkers},
  journal = {Sci. Rep.}, 
  volume  = {3},
  year    = {2013},
  pages   = {1467}
}

@ARTICLE{Chenetal2013,
  author  = {Chen, Q. and Wu, T. T. and Fang, M.},
  title   = {Detecting local community structure in complex networks based on local degree central nodes},
  journal = {Physica A.}, 
  volume  = {392},
  year    = {2013},
  pages   = {529-537}
}

@ARTICLE{Zhangetal2007,
  author  = {Zhang, S. and Wang, R. and Zhang, X.},
  title   = {Identification of overlapping community structure in complex networks using fuzzy c-means clustering},
  journal = {Physica A.}, 
  volume  = {374},
  year    = {2007},
  pages   = {483-490}
}

@ARTICLE{Nepuszetal2008,
  author  = {Nepusz, T. and Petr\'oczi, A. and N\'egyessy, L. and Bazs\'o, F.},
  title   = {Fuzzy communities and the concept of bridgeness in complex networks},
  journal = {Phys. Rev. E.}, 
  volume  = {77},
  year    = {2008},
  pages   = {016107}
}

@ARTICLE{FabricioLiang2013,
  author  = {Fabricio, B. and Liang, Z.},
  title   = {Fuzzy community structure detection by particle competition and cooperation},
  journal = {Soft Comput.}, 
  volume  = {17},
  year    = {2013},
  pages   = {659-673}
}

@ARTICLE{Sunetal2011,
  author  = {Sun, P. and Gao, L. and Han, S.},
  title   = {Identification of overlapping and non-overlapping community structure by fuzzy clustering in complex networks},
  journal = {Inf. Sci.}, 
  volume  = {181},
  year    = {2011},
  pages   = {1060-1071}
}

@ARTICLE{Wangetal2013,
  author  = {Wang, W. and Liu, D. and Liu, X. and Pan, L.},
  title   = {Fuzzy overlapping community detection based on local random walk and multidimensional scaling},
  journal = {Physica A.}, 
  volume  = {392},
  year    = {2013},
  pages   = {6578-6586}
}

@ARTICLE{Psorakisetal2011,
  author  = {Psorakis, I. and Roberts, S. and Ebden, M. and Sheldon, B.},
  title   = {Overlapping community detection using Bayesian non-negative matrix factorization},
  journal = {Phys. Rev. E.}, 
  volume  = {83},
  year    = {2011},
  pages   = {066114}
}

@CONFERENCE{ZhangYeung2012,
  author  = {Zhang, Y. and Yeung, D.},
  title   = {Overlapping Community Detection via Bounded Nonnegative Matrix Tri-Factorization},
  booktitle = {In Proc. ACM SIGKDD Conf.}, 
  year    = {2012},
  pages   = {606-614}
}

@ARTICLE{Liu2010,
  author  = {Liu, J.},
  title   = {Fuzzy modularity and fuzzy community structure in networks},
  journal = {Eur. Phys. J. B.}, 
  volume  = {77},
  year    = {2010},
  pages   = {547-557}
}

@ARTICLE{Havensetal2013,
  author  = {Havens, T. C. and Bezdek, J. C. and Leckie, C., Ramamohanarao, K. and Palaniswami, M.},
  title   = {A Soft Modularity Function For Detecting Fuzzy Communities in Social Networks},
  journal = {IEEE Trans. Fuzzy Syst.}, 
  volume  = {21},
  year    = {2013},
  pages   = {1170-1175}
}

@misc{Newman2013,
  author = {Newman, M. E. J.},
  title  = {Network data},
  howpublished = "\url{http://www-personal.umich.edu/~mejn/netdata/}",
  year = {2013}
}

@ARTICLE{SubeljBajec2012,
  author  = {\v{S}ubelj, L. and Bajec, M.},
  title   = {Ubiquitousness of link-density and link-pattern communities in real-world networks},
  journal = {Eur. Phys. J. B.}, 
  volume  = {85},
  year    = {2012},
  pages   = {1-11}
}

@ARTICLE{Lancichinettietal2008,
  author  = {Lancichinetti, A. and Fortunato, S. and Radicchi, F.},
  title   = {Benchmark graphs for testing community detection algorithms},
  journal = {Phys. Rev. E.}, 
  volume  = {78},
  year    = {2008},
  pages   = {046110}
}

@ARTICLE{Liuetal2014,
  author  = {Liu, W. and Pellegrini, M. and Wang, X.},
  title   = {Detecting Communities Based on Network Topology},
  journal = {Sci. Rep.}, 
  volume  = {4},
  year    = {2014},
  pages   = {5739}
}

@ARTICLE{Danonetal2005,
  author  = {Danon, L. and Diaz-Guilera, A. and Duch, J. and Arenas, A.},
  title   = {Comparing community structure identification},
  journal = {J. Stat. Mech.-Theory Exp.}, 
  volume  = {},
  year    = {2005},
  pages   = {P09008}
}

@ARTICLE{Gregory2011,
  author  = {Gregory, S.},
  title   = {Fuzzy overlapping communities in networks},
  journal = {J. Stat. Mech.-Theory Exp.}, 
  volume  = {},
  year    = {2011},
  pages   = {P02017}
}

@ARTICLE{LancichinettiFortunato2009,
  author  = {Lancichinetti, A. and Fortunato, S.},
  title   = {Benchmarks for testing community detection algorithms on directed and weighted graphs with overlapping communities},
  journal = {Phys. Rev. E.}, 
  volume  = {80},
  year    = {2009},
  pages   = {016118}
}

@CONFERENCE{HullermeierRifqi2009,
  author  = {Hullermeier, E. and Rifqi, M.},
  title   = {A Fuzzy Variant of the Rand Index for Comparing Clustering Structures},
  booktitle = {in Proc. IFSA/EUSFLAT Conf.}, 
  year    = {2009},
  pages   = {1294-1298}
}


%% file: references.bib
@ARTICLE{collins1975,
  title={Quantifier elimination for real closed fields by cylindrical algebraic decomposition},
  author={Collins, G. E.},
  journal={Lecture Notes in Computer Science},
  year={1975},
doi = "10.1007/3-540-07407-4\_17"
}

@book{basu2007,
  title     = {Algorithms in Real Algebraic Geometry},
  author    = {Basu, S. and Pollack, R. and Coste-Roy, M.F.},
  isbn      = {9783540330998},
  lccn      = {2006927110},
  series    = {Algorithms and Computation in Mathematics},
  year      = {2007},
  publisher = {Springer Berlin Heidelberg},
  doi = {10.1007/3-540-33099-2}
}

@book{bochnaketal1998,
  author    = {Bochnak, J. and Coste, M. and Roy, M-F.},
  publisher = {Springer},
  title     = {Real Algebraic Geometry},
  year      = {1998},
doi = {10.1007/978-3-662-03718-8}
}

@book{TRaAT,
author = {Baader, F. and Nipkow, T.},
title = {Term rewriting and all that},
year = {1998},
isbn = {0521455200},
publisher = {Cambridge University Press},
address = {USA},
doi = {10.1017/CBO9781139172752}
}

@phdthesis{locatelli,
  author  ={Locatelli, A.},
  title = {On the regularity of cylindrical algebraic decompositions},
  school = {University of Bath},
  year = {2016},
 type = {Ph{D} Thesis}
  }

@article{locatelli-paper,
author = {Davenport, J. H. and Locatelli, A. F. and Sankaran, G. K.},
title = {Regular cylindrical algebraic decomposition},
journal = {Journal of the London Mathematical Society},
volume = {101},
number = {1},
pages = {43-59},
doi = {https://doi.org/10.1112/jlms.12257},
year = {2020}
}

@phdthesis{wilson,
  author  ={Wilson, D.},
  title = {Advances in Cylindrical Algebraic Decomposition},
  school = {University of Bath},
  year = {2014},
type = {PhD Thesis}
  }

@ARTICLE{lazard2010,
  author       = {Lazard, D.},
  title        = {{CAD} and Topology of Semi-Algebraic Sets},
  journal      = {Math. Comput. Sci.},
  volume       = {4},
  number       = {1},
  pages        = {93--112},
  year         = {2010},
  doi          = {10.1007/S11786-010-0047-0}
}

@book{benedetti1990,
  title={Real Algebraic and Semi-algebraic Sets},
  author={Benedetti, R. and Risler, J.J.},
  isbn={9782705661441},
  lccn={90202291},
  series={Actualit{\'e}s math{\'e}matiques},
  year={1990},
  publisher={Hermann}
}

@article{schwartzsharir83,
title = {On the “piano movers” problem. {II}. {G}eneral techniques for computing topological properties of real algebraic manifolds},
journal = {Advances in Applied Mathematics},
volume = {4},
number = {3},
pages = {298-351},
year = {1983},
issn = {0196-8858},
doi = {https://doi.org/10.1016/0196-8858(83)90014-3},
author = {Jacob T Schwartz and Micha Sharir}
}

@book{dries_1998, 
place={Cambridge}, 
series={London Mathematical Society Lecture Note Series}, 
title={Tame Topology and O-minimal Structures}, 
DOI={10.1017/CBO9780511525919}, 
publisher={Cambridge University Press}, 
author={van den Dries, L.}, 
year={1998}, 
collection={London Mathematical Society Lecture Note Series}
}

@inproceedings{Strz17,
author = {Strzebo\'{n}ski, A.},
title = {CAD Adjacency Computation Using Validated Numerics},
year = {2017},
isbn = {9781450350648},
publisher = {ACM},
address = {New York, NY, USA},
doi = {10.1145/3087604.3087641},
booktitle = {Proceedings of the 2017 ACM on International Symposium on Symbolic and Algebraic Computation},
pages = {413–420},
numpages = {8},
location = {Kaiserslautern, Germany},
series = {ISSAC '17}
}

@inproceedings{LazardStyle23,
author = {Davenport, J. H. and Nair, A. S. and Sankaran, G. K. and Uncu, A. K.},
title = {Lazard-Style CAD and Equational Constraints},
year = {2023},
isbn = {9798400700392},
publisher = {Association for Computing Machinery},
address = {New York, NY, USA},
doi = {10.1145/3597066.3597090},
booktitle = {Proceedings of the 2023 International Symposium on Symbolic and Algebraic Computation},
pages = {218–226},
numpages = {9},
series = {ISSAC '23}
}

@InProceedings{McCallum98,
author="McCallum, S.",
editor="Caviness, Bob F.
and Johnson, Jeremy R.",
title="An Improved Projection Operation for Cylindrical Algebraic Decomposition",
booktitle="Quantifier Elimination and Cylindrical Algebraic Decomposition",
year="1998",
publisher="Springer Vienna",
address="Vienna",
pages="242--268",
isbn="978-3-7091-9459-1",
doi = "10.1007/978-3-7091-9459-1 \_12"
}

@ARTICLE{Brown01,
title = {Improved Projection for Cylindrical Algebraic Decomposition},
journal = {Journal of Symbolic Computation},
volume = {32},
number = {5},
pages = {447-465},
year = {2001},
issn = {0747-7171},
author = {Brown, C. W.},
doi = {10.1006/jsco.2001.0463}
}

@ARTICLE{CAC21,
title = {Deciding the consistency of non-linear real arithmetic constraints with a conflict driven search using cylindrical algebraic coverings},
journal = {Journal of Logical and Algebraic Methods in Programming},
volume = {119},
pages = {100633},
year = {2021},
issn = {2352-2208},
author = {Ábrahám, E. and Davenport, J. H. and England, M. and Kremer, G.}
}

@ARTICLE{Jasper,
title = {Levelwise construction of a single cylindrical algebraic cell},
journal = {Journal of Symbolic Computation},
volume = {123},
pages = {102288},
year = {2024},
issn = {0747-7171},
author = {Jasper Nalbach and Erika Ábrahám and Philippe Specht and Christopher W. Brown and James H. Davenport and Matthew England}
}

@inproceedings{RegularChains,
author = {Chen, C. and Maza, M. M.},
title = {Quantifier Elimination by Cylindrical Algebraic Decomposition Based on Regular Chains},
year = {2014},
isbn = {9781450325011},
publisher = {ACM},
address = {New York, NY, USA},
doi = {10.1145/2608628.2608666},
booktitle = {Proceedings of the 39th International Symposium on Symbolic and Algebraic Computation},
pages = {91–98},
numpages = {8},
location = {Kobe, Japan},
series = {ISSAC '14}
}

@inproceedings{NuCAD,
author = {Brown, C. W.},
title = {Projection and Quantifier Elimination Using Non-Uniform Cylindrical Algebraic Decomposition},
year = {2017},
isbn = {9781450350648},
publisher = {ACM},
address = {New York, NY, USA},
doi = {10.1145/3087604.3087651},
booktitle = {Proceedings of the 2017 ACM on International Symposium on Symbolic and Algebraic Computation},
pages = {53–60},
numpages = {8},
location = {Kaiserslautern, Germany},
series = {ISSAC '17}
}

@InProceedings{ComplexityCAD,
author="England, M.
and Davenport, J. H.",
editor="Gerdt, Vladimir P.
and Koepf, Wolfram
and Seiler, Werner M.
and Vorozhtsov, Evgenii V.",
title="The Complexity of Cylindrical Algebraic Decomposition with Respect to Polynomial Degree",
booktitle="Computer Algebra in Scientific Computing",
year="2016",
publisher="Springer International Publishing",
address="Cham",
pages="172--192",
isbn="978-3-319-45641-6"
}

@InProceedings{MLCAD,
author="Huang, Z.
and England, M.
and Wilson, D.
and Davenport, J. H.
and Paulson, L. C.
and Bridge, J.",
title="Applying Machine Learning to the Problem of Choosing a Heuristic to Select the Variable Ordering for Cylindrical Algebraic Decomposition",
booktitle="Intelligent Computer Mathematics",
year="2014",
publisher="Springer International Publishing",
address="Cham",
pages="92--107",
isbn="978-3-319-08434-3",
doi = "10.1007/978-3-319-08434-3 \_8"
}

@ARTICLE{BrownSimple,
title = {Simple {CAD} {C}onstruction and its {A}pplications},
journal = {Journal of Symbolic Computation},
volume = {31},
number = {5},
pages = {521-547},
year = {2001},
issn = {0747-7171},
doi = {https://doi.org/10.1006/jsco.2000.0394},
author = {C. W. Brown}
}

@InProceedings{Arnon,
author="Arnon, D. S.
and Collins, G. E.
and McCallum, S.",
editor="Caviness, B. F.
and Johnson, J. R.",
title="Cylindrical {A}lgebraic {D}ecomposition {I}: {T}he {B}asic {A}lgorithm",
booktitle="Quantifier Elimination and Cylindrical Algebraic Decomposition",
year="1998",
publisher="Springer Vienna",
address="Vienna",
pages="136--151",
isbn="978-3-7091-9459-1",
doi = "10.1007/978-3-7091-9459-1 \_6"
}

@Article{vDDM,
 Author = {van den Dries, Lou and Miller, Chris},
 Title = {Geometric categories and o-minimal structures},
 FJournal = {Duke Mathematical Journal},
 Journal = {Duke Math. J.},
 ISSN = {0012-7094},
 Volume = {84},
 Number = {2},
 Pages = {497--540},
 Year = {1996},
 Language = {English},
 DOI = {10.1215/S0012-7094-96-08416-1},
}

@article{TR1972,
author = {Aho, A. V. and Garey, M. R. and Ullman, J. D.},
title = {The Transitive Reduction of a Directed Graph},
journal = {SIAM Journal on Computing},
volume = {1},
number = {2},
pages = {131-137},
year = {1972},
doi = {10.1137/0201008},
}

@inproceedings{miniCAD,
author = {Michel, Lucas and Mathonet, Pierre and Zena\"{\i}di, Na\"{\i}m},
title = {On Minimal and Minimum Cylindrical Algebraic Decompositions},
year = {2024},
isbn = {9798400706967},
publisher = {Association for Computing Machinery},
address = {New York, NY, USA},
doi = {10.1145/3666000.3669704},
booktitle = {Proceedings of the 2024 International Symposium on Symbolic and Algebraic Computation},
pages = {316–323},
numpages = {8},
location = {Raleigh, NC, USA},
series = {ISSAC '24}
}
